\documentclass[usenatbib]{mn2e}
\usepackage{epsfig}
\usepackage{subfigure}
\usepackage{amssymb}
\usepackage{amsmath}
\usepackage{color}
\usepackage{float}

\newcommand{\bn}{\begin{enumerate}}
\newcommand{\en}{\end{enumerate}}
\newcommand{\bi}{\begin{itemize}}
\newcommand{\ei}{\end{itemize}}

\def\gtorder{\mathrel{\raise.3ex\hbox{$>$}\mkern-14mu
    \lower0.6ex\hbox{$\sim$}}} 
\def\ltorder{\mathrel{\raise.3ex\hbox{$<$}\mkern-14mu
    \lower0.6ex\hbox{$\sim$}}}

\newcommand{\apj}{ApJ}

\newcommand{\apjl}{ApJL}
\newcommand{\apjs}{ApJS}
\newcommand{\mnras}{MNRAS}
\newcommand{\aj}{AJ}

\newcommand{\pasj}{PASJ}

\newcommand{\na}{New Astronomy}

\title[Direct Collapse to Supermassive Black Hole Seeds]
{Direct Collapse to Supermassive Black Hole Seeds with Radiation Transfer: Cosmological Halos}

\author[Ardaneh et al.]
{Kazem Ardaneh$^{1}$\thanks{E-mail: kazem.arrdaneh@gmail.com},
Yang Luo$^{1,2}$, Isaac Shlosman$^{1,2}$\thanks{E-mail: shlosman@pa.uky.edu}, Kentaro Nagamine$^{1,3,4}$, 
\newauthor John H. Wise$^{5}$, Mitchell C. Begelman$^{6}$
\\
$^{1}$ Theoretical Astrophysics, Department of Earth \& Space Science, Graduate School of Science, Osaka University, 
   Osaka 560-0043, Japan\\
$^{2}$ Department of Physics \& Astronomy, University of Kentucky, Lexington, KY 40506-0055, USA\\
$^{3}$ Department of Physics \& Astronomy, University of Nevada, Las Vegas, NV 89154-4002, USA\\
$^{4}$ Kavli IPMU, The University of Tokyo, 5-1-5 Kashiwanoha, Kashiwa, Chiba, 277-8583, Japan\\
$^{5}$ Center for Relativistic Astrophysics, Georgia Institute of Technology, Atlanta, GA 30332, USA\\
$^{6}$ JILA, University of Colorado and National Institute of Standards and Technology, 440 UCB, Boulder, 
    CO 80309-0440, USA\\
}

\begin{document}
 
\date{Accepted 2018 June 17; Received 2018 June 8; in original form 2018 March 7}

%\pagerange{\pageref{firstpage}--\pageref{lastpage}} \pubyear{2014}

\maketitle

\begin{abstract}
We have modeled direct collapse of a primordial gas within dark matter halos in the presence of radiative transfer,
in high-resolution zoom-in simulations in a cosmological framework, down to the formation of the photosphere and 
the central object. Radiative transfer has
been implemented in the flux-limited diffusion (FLD) approximation. Adiabatic models were run for 
comparison. We find that ($a$) the FLD flow forms an irregular central structure  
and does not exhibit fragmentation, contrary to adiabatic flow which forms a thick disk, driving a pair 
of spiral shocks, subject to Kelvin-Helmholtz shear instability forming fragments; ($b$) the growing central core in 
the FLD flow quickly reaches $\sim 10\,M_\odot$ and a highly variable luminosity of $10^{38}-10^{39}\,{\rm
erg\,s^{-1}}$, comparable to the Eddington luminosity. It experiences massive recurrent outflows driven 
by radiation force and thermal pressure gradients, which mix with the accretion
flow and transfer the angular momentum outwards; and ($c$) the interplay between these processes
and a massive accretion, results in photosphere at $\sim 10$\,AU.
We conclude that in the FLD model (1) the central object exhibits dynamically insignificant rotation
and slower than adiabatic temperature rise with density;
(2) does not experience fragmentation leading to star formation, thus promoting
the fast track formation of a supermassive black hole (SMBH) seed; (3) inclusion of radiation force 
leads to outflows, resulting in the mass accumulation within the central $10^{-3}$ pc, which is  
$\sim$100 times larger than characteristic scale of star formation.
The inclusion of radiative transfer reveals complex early stages 
of formation and growth of the central structure in the direct collapse scenario of SMBH seed formation.
\end{abstract}


%%%%%%%%%%%%%%%%%%%%%%%%%
\section{Introduction}
\label{sec:intro}

Supermassive black holes (SMBHs) of $\gtorder 10^9\,M_\odot$ are increasingly found at high redshifts, 
$z\gtorder 6$ \citep[e.g.,][]{fan03,will10,mor11,wu15}, up to $z\sim 7.54$ at present 
\citep[e.g.,][]{vene17,bana18}. They reside in very luminous active galactic nuclei (AGN) and appear to 
form the extreme of the overall population of such objects. 

While 2 or 3 specific alternatives have been explored for explaining the
development of such massive SMBHs at these redshifts, the broader
issue of SMBH formation is equally important for our understanding of structure evolution
in the universe and galaxy evolution \citep[e.g.,][for review]{shlo13}. 
The main difficulty in forming the SMBHs in galaxy centers is their long growth time, if the 
initial seed mass is small compared to the final product. At high $z$, the only realistic options include
direct collapse leading to massive seeds \citep[e.g.,][]{hae93,bromm03,beg06,wise08,beg09,mil09,reg09,sch10,
hos11,choi13,choi15,lat13a,lat13b,shlo16}, or remnants of Population\,III stars, which, if fed vigorously
and merged, can form less massive SMBHs of Seyfert galaxies, in principle. 

Conditions for direct collapse with a gas of pristine composition are relatively simple. Namely, if the gas
can form molecular hydrogen and cool below $\sim 1000$\,K, it will collapse into DM minihalos of $\sim 
10^6\,M_\odot$, while if the gas remains atomic, it requires DM halos with virial temperatures reaching 
$\sim 10^4$\,K. In the former case, the process is expected to lead to the formation of a Pop\,III star 
or a few 
stars \citep[e.g.,][]{abel02,bromm02,oshea07,turk09}. In the presence of radiative feedback, a rather 
normal initial mass function (IMF) has been found \citep[e.g.,][]{hos11,hir15,hos16}.
In the latter case, the result can be the formation of a supermassive star \citep[SMS,][]{beg06,beg10}, which 
evolves to a SMBH seed after its core collapse --- a so-called quasi-star which accretes in the hyper-Eddington 
regime. Supersonic streaming velocities --- remnants from the recombination epoch --- 
can suppress formation of Pop\,III stars, resulting in more massive central objects \citep{tan13,hir17}.

The quasi-star mass has been argued to lie in the range of $\sim 10^{1}-10^{6}\,M_\odot$, or even higher
\citep{beg08}, if fragmentation can be 
suppressed and the angular momentum can be efficiently transferred outward \citep{beg09,beg10,choi13,choi15}. 
Gravitational torques assisted by shocks have been verified to transfer the angular momentum of the collapsing 
gas to the DM and the outer gas \citep{choi13,choi15}. 
Supersonic turbulent motions, induced by the torques and resulting shear, damp fragmentation in the atomic gas.
This evolution differs from that described by the self-similar analysis, which was necessarily limited to a 
linear stage \citep{hana00}; the growing bar-like $m=2$ mode in its nonlinear stage does not lead to 
fragmentation, but induces gas inflow.
Alternatively, it is possible that the stellar evolution 
stage can be by-passed completely, e.g., if the temperature does not rise sufficiently high to trigger
thermonuclear reactions \citep[e.g.,][]{beg09,choi13,shlo16}. 

Direct collapse within dark matter (DM) halos has been investigated in the optically-thin regime, allowing
the gas to cool down efficiently. Difficulties in performing on-the-fly 
radiative transfer in the collapsing 
gas have motivated models that switch to an adiabatic equation of state, by cutting off cooling
within a region having a density above some critical value \citep[e.g.,][]{bec15,bec17}. These models lead  
to a rotationally
dominated flow in the inner region --- a disk and fragmentation due to Jeans instability. 
However, this approximation
neglects a long list of processes that operate in the central region, such as radiation pressure and its force,
and artificially suppresses cooling in regions where it should be able to operate.

In a recent paper \citep{luo18}, we have dealt with the optically-thick stage of direct collapse within
isolated DM halos, including radiative transfer in the flux-limited diffusion (FLD) approximation.
Furthermore, we have compared the FLD models with models assuming 
an adiabatic equation of state, and found that their evolutions diverge. The main goal of current work is
to model the optically-thick stage of direct collapse with radiation transfer in the FLD regime, in a fully
self-consistent cosmological framework. 
 
The formation of the central object under direct collapse has not been simulated 
so far, except under simplified adiabatic conditions, 
with the exception of \citet{luo18} which dealt with 
isolated DM halos. In this paper, we take an additional step towards realistic modeling of the outcome.
If the collapse leads to an SMS, we wish to determine its properties, namely, its mass, size and
spin. What opacities dominate in its interior and near its photosphere? If it is rotation-dominated, does it 
rotate differentially, or more like a solid body? How does the massive accretion flow 
affect its internal structure, including convection, and its thermal and
dynamical relaxation? What is its central temperature and is it sufficiently high to trigger thermonuclear
reactions? Are outflows associated with this stage and what drives them? Inclusion of radiation transfer
should in principle answer these questions.

This paper is structured as follows. The next section describes the numerical code used in our simulation, as well
as the details of radiative transfer implemented here, and the initial cosmological conditions used.
Section\,3 presents our results for adiabatic flows, and Section\,4 for FLD flows. This is followed
by a discussion Section and conclusions of this work. 
 
\section{Numerical techniques}
\label{sec:method}

For simulations of direct collapse within DM halos, we invoke the modified Eulerian adaptive mesh refinement (AMR) code 
Enzo-2.4 \citep{bryan97,nor99}. Enzo implements a particle-mesh $N$-body method to calculate the
gravitational dynamics, to follow collisionless DM particles, and a second-order piecewise parabolic method
to solve hydrodynamics \citep[][]{cole84,bryan95}. Supplementary inner meshes are allowed by the structured AMR,
in order to enhance the resolution in the prespecified regions. The number of
rectangular grids that cover some region at a given refinement level, and the number of refinement levels
are not subject to limitation \citep{ber89}. When densities in the DM or gas exceed $\rho_0 N^l$, the simulation
grid is refined  by a factor of two in length scale, where $N = 2$ is the refinement factor, $l$ is the maximal AMR
level of refinement, and $\rho_0$ is the threshold density for refinement. The force resolution  corresponds to
twice the minimal cell size in adaptive PM codes \citep[e.g.,][]{kra97}

The \citet{true97} condition to resolve the Jeans length, i.e., to have at least four cells per Jeans length, 
is exceeded, in order to avoid spurious fragmentation. We have imposed the condition
of at least 8 cells per Jeans length in our simulations, in line with recent numerical experiments requiring even higher
numerical resolution in order to properly resolve the turbulent motions \citep[e.g.,][]{sur10,fede11,turk12,lat13a}.

%%%%%%%%%%%%%%%%%%%%%%%%%%%%%%%%%%%%%%%%%%%
\subsection{Radiation Hydrodynamics and Radiative Transfer Formalism}
\label{sec:rad-transf}
%%%%%%%%%%%%%%%%%%%%%%%%%%%%%%%%%%%%%%%%%%%

We use the flux-limited diffusion (FLD) approximation in order to model radiation transport.
Local thermodynamic equilibrium (LTE) is imposed in optically-thick regions of the flow, where the Planck 
intensity is used for the gas emissivity, and the  Saha equation determines the gas ionization level  
\citep[e.g.,][]{ryb79}.

In Enzo-2.4, in order to solve the radiation transport equation, a fully implicit inexact Newton method has been 
applied
and coupled to the AMR cosmological hydro solver with an explicit, operator-split algorithm, but only at the end
of the top level timestep \citep{rey09}. We have  modified this by updating each level of refinement at the end 
of respective timestep. This makes the FLD fully consistent with the hydrodynamics \citep{luo18}.

In \citet{luo18} and here, we have introduced  the radiation force and $v/c$ order terms,
with $c$ and $v$ being the speed of light and the gas velocity, respectively. In the current work, we have included
the cosmological terms, with $a$ being the cosmological expansion parameter. The modified Euler equation is 

\begin{equation}
\label{eq:euler}
\frac{\partial \rho \mathbf{v}}{\partial t}  
    + \frac{1}{a} \nabla\cdot (\rho \mathbf{v} \mathbf{v} + \mathbb{I} p )  = 
    - \frac{\dot{a}}{a} \rho \mathbf{v} - \frac{1}{a}\rho\nabla \phi + \frac{1}{a}\frac{\kappa_{\rm R}}{c}{\bf F},
\end{equation}

Here, $p$, $\rho$ and  ${\bf F}$ are the thermal pressure, comoving baryon density and  the radiation energy flux, 
respectively.  $\mathbb{I}$ is the identity matrix,  and $\kappa_{\rm R}$ is the Rosseland mean opacity 
(\S\,\ref{sec:opacity}). The gravitational potential $\phi$ is calculated from the DM density $\rho_{\rm DM}$ and
the baryon density $\rho$. 

The radiative flux vector in the FLD approximation can be presented by the gradient of radiation energy 
density \citep{lev84}, i.e., using the  Fick's diffusion law, 

\begin{equation}
\label{eq:Fick}
 {\bf F} = - \frac{c\lambda_{\rm F}}{\kappa_{\rm R}}\nabla E,
\end{equation}
where $\lambda_{\rm F}$ is the {\it flux limiter}, $\lambda_{\rm F} = \lambda_{\rm F}(E, \nabla E 
\kappa_{\rm R}) = (9 + R_\lambda^2)^{-1/2}$. The auxiliary function $R_\lambda$ is defined as    
$R_\lambda={|\nabla E|}/({\kappa_{\rm R} E})$.  

The comoving radiation energy density in a cosmological medium,
omitting the frequency-dependence by integration over the radiation energy spectrum, is given by 

\begin{equation}
\begin{aligned}\label{eq:radiation_PDE}
  \frac{\partial E}{\partial t} + \frac{1}{a} &\nabla\cdot (E {\bf v}) = \\
  & - \frac{1}{a^2}\nabla\cdot {\bf F} - \frac{\dot{a}}{a} E - \mathbb{P}:\nabla {\bf v}- c\kappa_{\rm P} E 
    + \eta  - \frac{1}{a}\frac{\kappa_{\rm R}}{c}{\bf F}\cdot \mathbf{v}
\end{aligned}    
\end{equation}
\citep{rey09,ENZO14}.  Here parameter $\eta$ is the blackbody emissivity given by 
$\eta = 4\kappa_{\rm P}\,\sigma_{\rm SB}\,T^4$, 
where  $T$ is the gas temperature,  $\sigma_{\rm SB}$ is the Stefan-Boltzmann constant, and  $\kappa_{\rm P}$ 
is the Planck mean opacity.
The term has been added to the above equation, and the radiation pressure tensor, $\mathbb{P}$, 
has been written using auxiliary functions,

\begin{equation}
\begin{aligned}
\label{eq:prad}
  &\mathbb{P} = \mathbb{D}E\\ 
  &\mathbb{D}=\frac{1-\chi}{2}\mathbb{I}+\frac{3\chi-1}{2}\mathbf{n}\otimes \mathbf{n}\\ 
  &\chi=\lambda_{\rm F}+\lambda_{\rm F}^2R^2\\
  &\mathbf{n}=\frac{\nabla E}{|\nabla E|}.
\end{aligned}
\end{equation}

The equation of the gas energy density $e$ has been modified by introducing the $O(v/c)$ term,

\begin{equation}
\begin{aligned}
 \frac{\partial e}{\partial t} + \frac{1}{a} & \nabla\cdot [(e+p)\mathbf{v}]  = \\
    &- \frac{2\dot{a}}{a} e - \frac{1}{a}\rho\mathbf{v}\cdot\nabla\phi 
    + c \kappa_{\rm P} E - \eta + \frac{1}{a}\frac{\kappa_{\rm R}}{c}{\bf F}\cdot\mathbf{v},
    \label{eq:total_energy} 
\end{aligned}
\end{equation}

%%%%%%%%%%%%%%%%%%%%%%%%%%%%%%%%%%%%%%%%
\subsection{Opacities}
\label{sec:opacity}
%%%%%%%%%%%%%%%%%%%%%%%%%%%%%%%%%%%%%%%%

\begin{figure}
\begin{center}
\includegraphics[width=0.5\textwidth,angle=0] {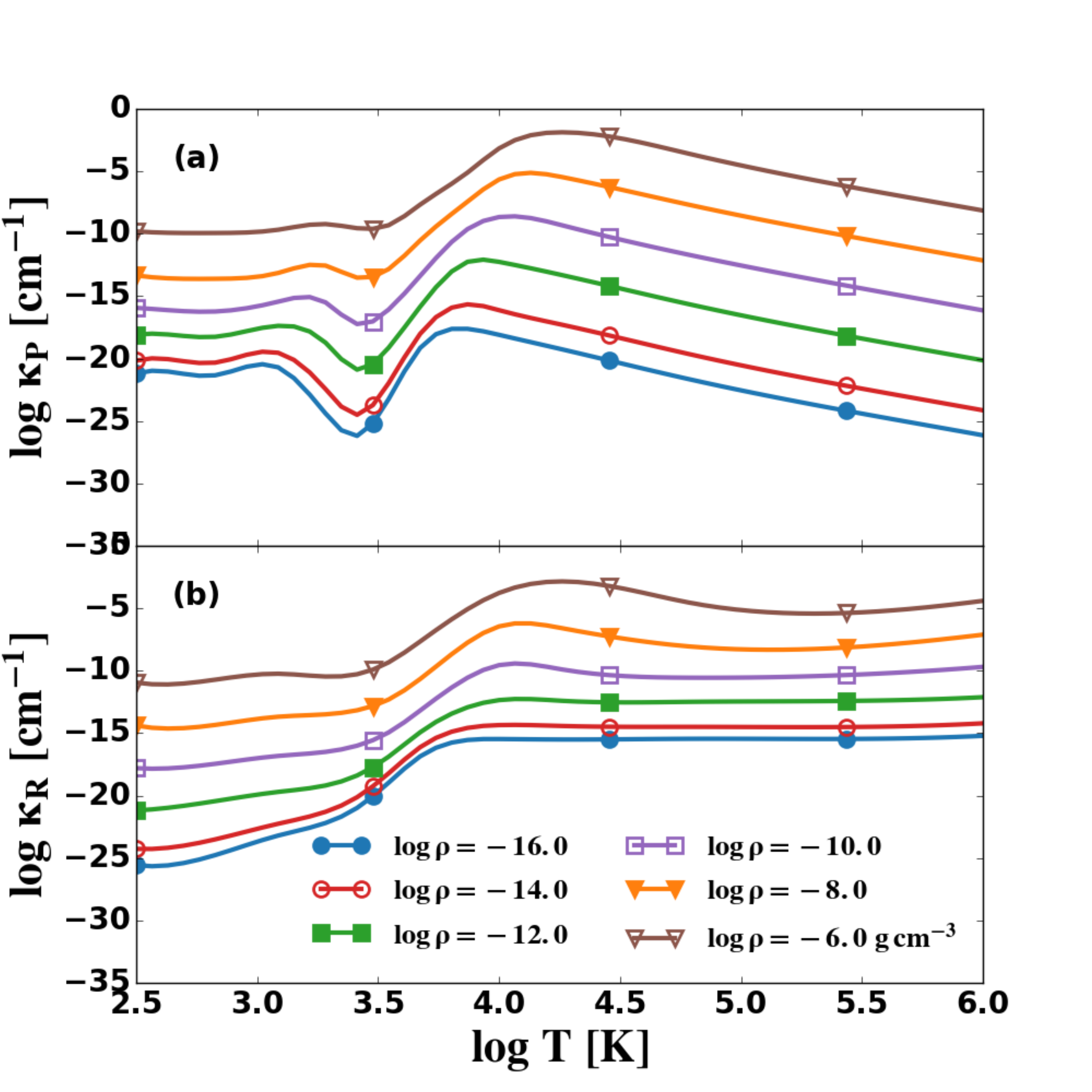}
\caption{{\it (a)} Planck mean and {\it (b)} Rosseland mean opacities as a function of temperature for different 
densities, adopted from \citet{may05}. These opacities are defined as area per unit volume, and hence have
units per unit length.
The opacity table in \citet{may05} covers a density range from $-16 < \log\rho\ {\rm [g\,cm^{-3}]} < -2$ and a 
temperature range of $1.8 <\,\log\,T\,{\rm [K]}\,< 4.5$. The table has been extrapolated for temperatures 
above $2\times10^4\ {\rm K}$.}
\label{opacity}
\end{center}
\end{figure}

We use tabulated opacities \citep{may05} in the form of Planck and Rosseland mean opacities for
matter with a primordial composition (Fig.\,\ref{opacity}). Three elements have been included, namely, 
H, He, and Li. The following H species have been accounted for: H, H$^-$, H$^+$, H$_2$, H$_2^+$, H$_3^+$, 
as well as D, He and Li.

The tables cover  the ranges $-16 < {\rm log}\,\rho\, {\rm [g\,cm^{-3}]} < -2$ in density, and 
$1.8 < {\rm log}\,T\, {\rm [K]} < 4.5$ in temperature (Fig.\,\ref{opacity}). The 
temperature-dependence of the opacity has been extrapolated by using the analytic expressions for the free-free, 
bound-free and electron-scattering opacities.

%% FIGURE 1

%%%%%%%%%%%%%%%%%%%%%%%%%%%%%%%%%%%%%%%%%%%%%%
\subsection{Cooling and heating rates}
\label{sec:cool-heat}
%%%%%%%%%%%%%%%%%%%%%%%%%%%%%%%%%%%%%%%%%%%%%%

Cooling rates from \citet{luo16} have been adopted for the optically-thin part of the collapse, 
solving for internal energy and  radiative cooling. A non-equilibrium primordial chemistry network 
has been invoked to calculate  pressure, temperature, ratio of specific heats, and cooling time,
for  atomic H and He \citep[][]{abel97,ann97}. For this purpose, we used  GRACKLE 1.1
package \citep[][https://grackle.readthedocs.org/]{ENZO14,kim14,smi17}.
We have assumed dust-free gas, and calculated radiative cooling and heating rates accounting
for  collisional excitation and free-free transitions, recombination, and atomic line excitation. 

For comparison models with  adiabatic equation of state, we have used  the
local Jeans length, $\lambda_{\rm J}$, for each cell, i.e., $\tau=\kappa_{\rm P}\lambda_{\rm J}$, to
calculate the optical depth. An exponential cutoff in the cooling rate has been imposed for the
optically-thin-to-thick transition region,
\begin{equation}
\label{eq:CoolCutoff}
  \Lambda = \Lambda_{\rm thin}{\rm e}^{-\tau},
\end{equation}
where $\Lambda_{\rm thin}$ is the optically-thin cooling rate.

LTE has been assumed for the optically-thick part of the adiabatic and FLD flows. 
A number of  alternative options have been used to determine the position of the
photosphere in the FLD flow (Section\,\ref{sec:photo}).

%%%%%%%%%%%%%%%%%%%%%%%%%%%%%%%%%%%%%%%%%%%%%
\subsection{Cosmological initial conditions}
\label{sec:ICs}
%%%%%%%%%%%%%%%%%%%%%%%%%%%%%%%%%%%%%%%%%%%%%

We use fully cosmological initial conditions (ICs) for our current runs and invoke zoom-in simulations  
\citep[e.g.,][]{choi15,shlo16,luo16}.
The gas density exceeds the DM density on spatial scales smaller than $\sim 0.3-3$\,pc, 
and hence the gas decouples from the DM.  
For the gas, the gravitational softening is adaptive and varies with refinement level.

While the cosmological evolution is of course tied to the time since the Big Bang, we find it helpful
to renormalize the time in our simulations to $t=0$ when the optical depth in the collapsing gas just 
exceeds unity. Time before this benchmark is considered negative.
 
We use the WMAP5 cosmology \citep{kom09}. $\Omega_{\Lambda}= 0.721$, $\Omega_{\rm m} = 0.279$,  
$\Omega_{\rm b} = 0.0445$, $h = 0.701$, $\sigma_{8}= 0.807$, and $n_{\rm s} = 0.961$. The cosmological 
ICs are initialized at $z = 199$ with the MUSIC code \citep{hahn11}, as described in \citet{luo16}.

Generating a set of zoom-in ICs is a two-step process. First, we produce $1\,h^{-1}$\,Mpc comoving $128^3$
DM-only ICs for the pathfinder simulation, and run it without AMR until $z=10$. Using the HOP
group finder \citep{eis98}, we select an appropriate DM halo whose mass is $\gtorder 10^8\,h^{-1}\,{\rm 
M_\odot}$ at $z=10$. Secondly, we generate $0.18\,h^{-1}$\,Mpc ICs with $512^3$ effective resolution in DM 
and gas embedded in the zoom-in region. Since we use the
same random seeds for these ICs as the first step, the phases of both
ICs are identical. The zoom-in region is centered on the selected halo
position and is set to be large enough to cover the Lagrangian volume of the selected halo and the 
immediate neighborhood.

%% FIGURE 2
\begin{figure*}
\centerline{
 \includegraphics[width=0.87\textwidth,angle=0] {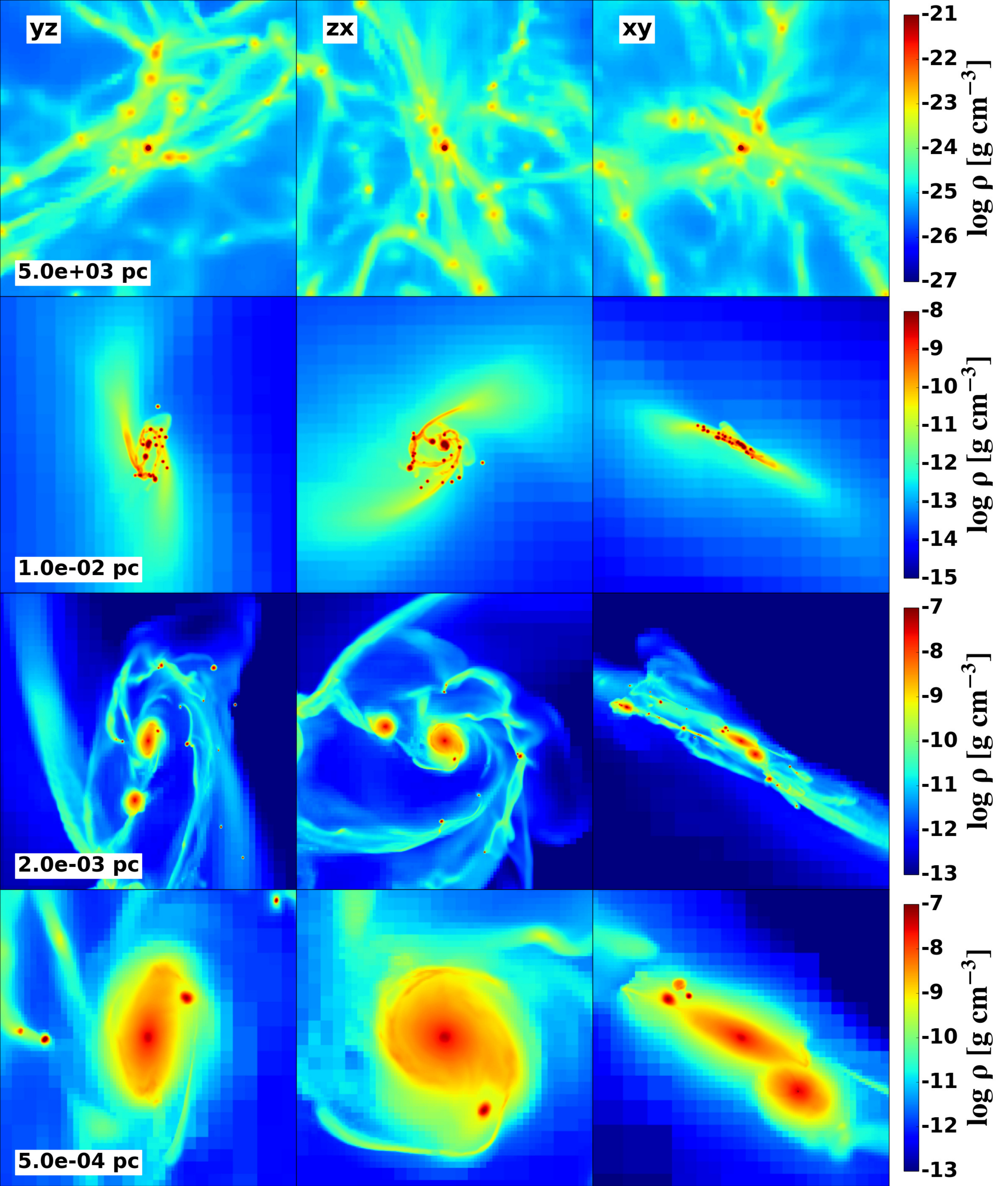}
}
\caption{Adiabatic collapse: density-weighted map of the gas density projected onto the $yz$-plane
(left column), $zx$-plane (middle column), and $xy$-plane (right column), on four spatial scales:
5\,kpc (top row), $10^{-2}$\,pc (2nd row), $2\times 10^{-3}$\,pc (3rd row), and $5\times 
10^{-4}$\,pc (bottom row). All panels shown at the end of the simulation at $t=220.8$\,yr.
}
\label{fig:adia_proj_end}
\end{figure*} 

We have measured the spin parameters of DM halos within the relevant mass and cosmological spin ranges,
$\lambda\sim 0.01-0.07$, in the zoom region at $z\sim 10$. Out of this range, we chose a DM halo with 
$\lambda\sim 0.05$, which is close to the average spin \citep[e.g.,][]{bull01,col18}. The gravitational collapse
happens at $z\sim 15.8$ in both the adiabatic and FLD runs. At this time, the DM halo has a virial mass
of $1.4\times 10^7\,h^{-1}$M$_\odot$ and virial radius of $\sim 0.5\,h^{-1}$\,kpc.

%% FIGURE 3
\begin{figure*}
\centerline{
 \includegraphics[width=0.92\textwidth,angle=0] {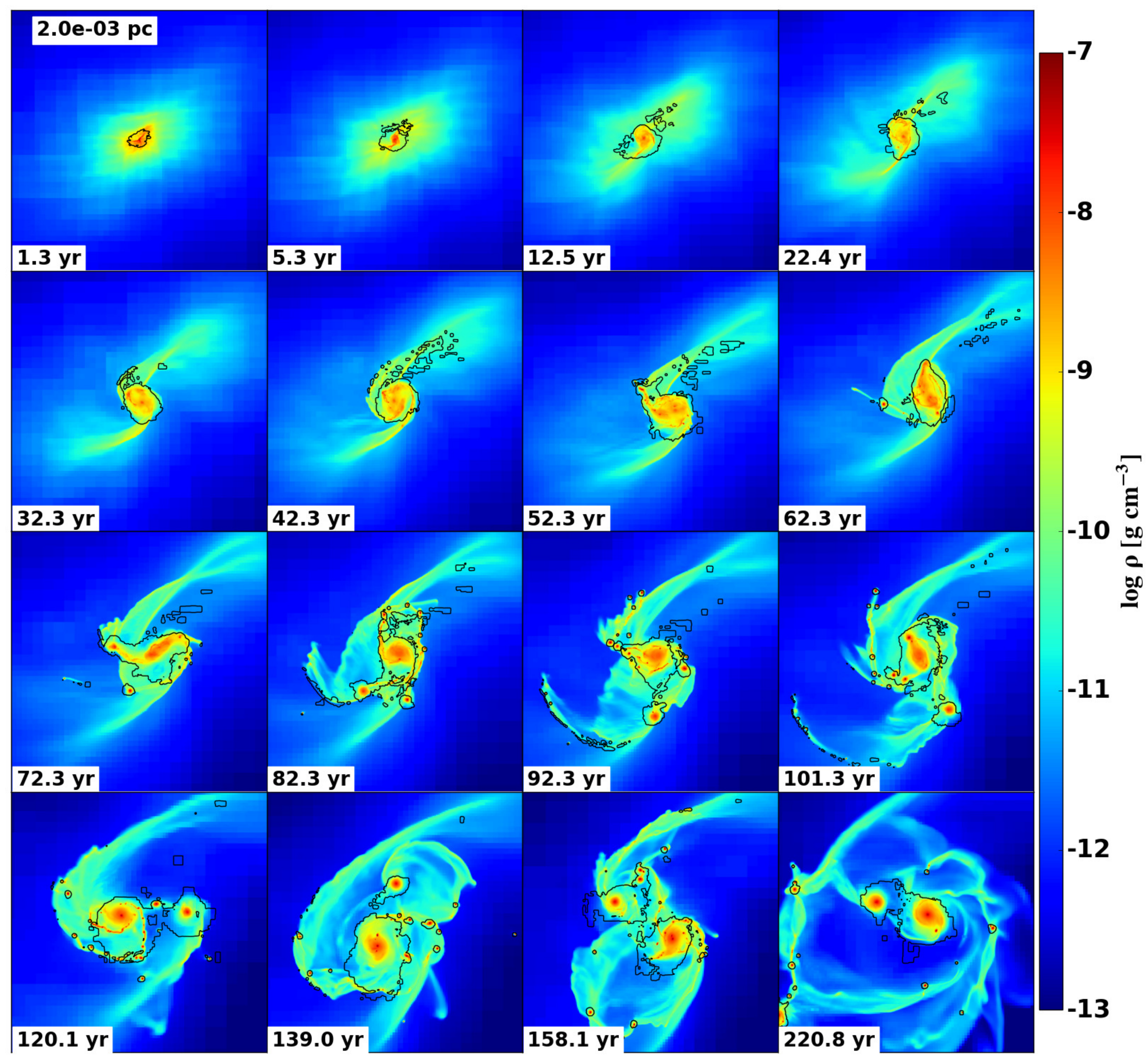}
}
\caption{Adiabatic collapse: evolution of the gas density-weighted  projection on the $zx$-plane
(face-on view along the $y$-axis) for the central $2\times 10^{-3}$\,pc. The black continuous
contour represents the position of the `photosphere', at $\tau=1$ (see Section\,\ref{sec:cool-heat}). The 
time $t=0$ is defined by the appearance of the photosphere. 
}
\label{fig:adia_proj}
\end{figure*}

%% FIGURE 4
\begin{figure*}
\centerline{
 \includegraphics[width=0.77\textwidth,angle=0] {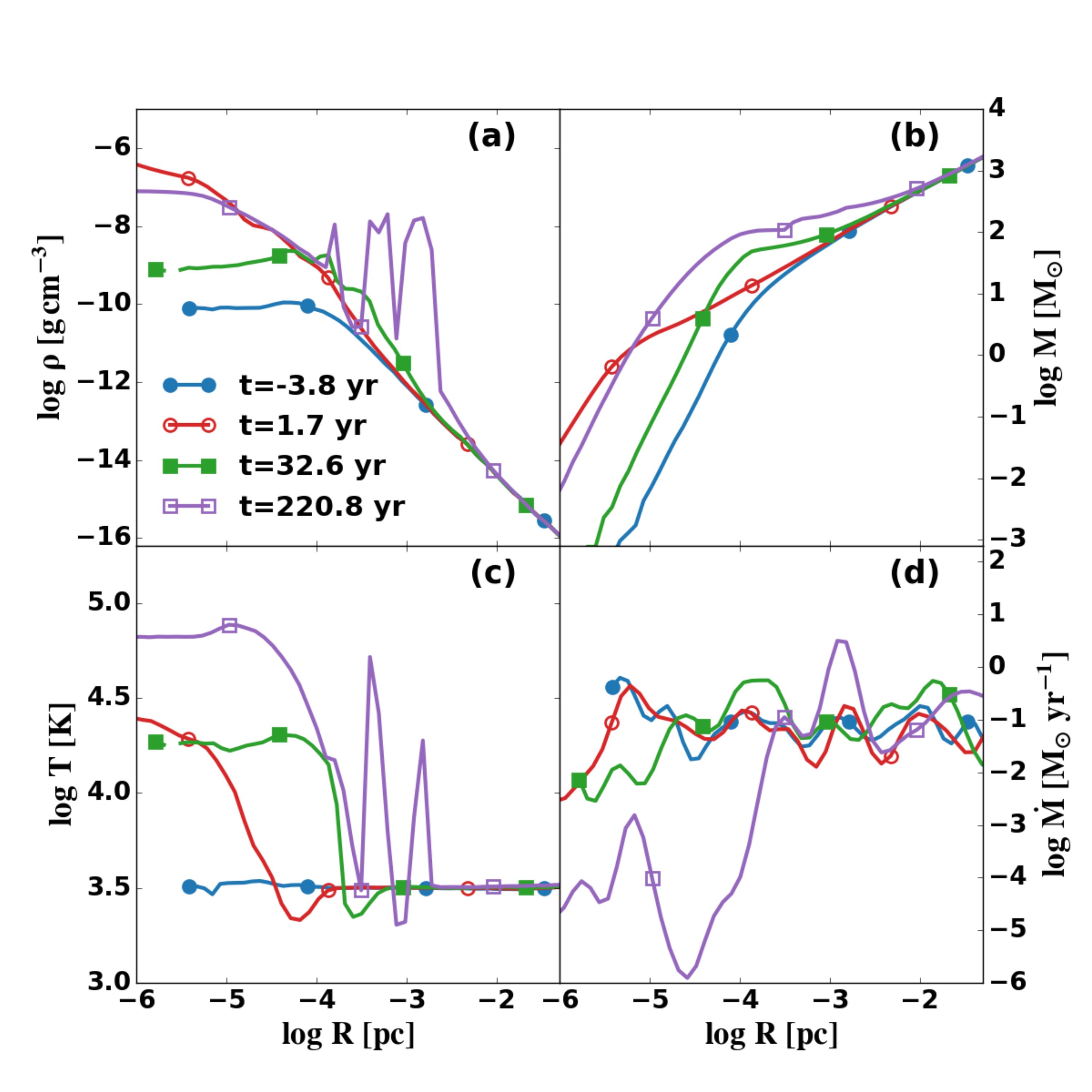}
}
\caption{Adiabatic collapse: evolution of basic parameters of the flow at few representative times: 
{\it (a)} mass density profiles;
{\it (b)} mass within a fixed spherical radius; {\it (c)} temperature profiles; {\it (d)} mass accretion 
rate profiles.
}
\label{fig:adia_param} 
\end{figure*} 

%%%%%%%%%%%%%%%%%%%%%%%%%%%%%%%%%%
\section{Results: Adiabatic flow}
\label{sec:adia_results}
%%%%%%%%%%%%%%%%%%%%%%%%%%%%%%%%%% 
 
%% FIGURE 5
\begin{figure*}
\centerline{
 \includegraphics[width=0.92\textwidth,angle=0] {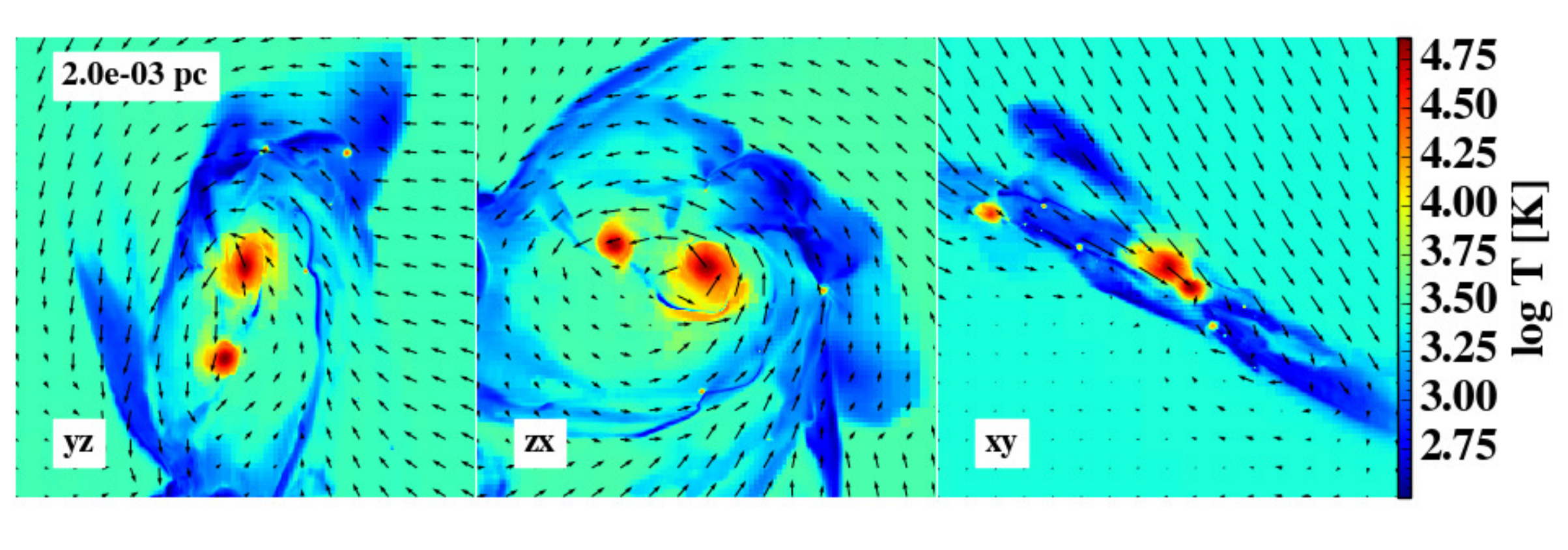}
}
\caption{Adiabatic collapse: density-weighted map of the temperature projected onto the three principal
planes at the end of the simulation at $t=220.8$\,yr, positioned on the center of mass. Overplotted arrows 
represent the velocity field and their length is proportional to the velocity magnitude (see Figure\,
\ref{fig:adia_vels}). The colors represent the gas temperature.
}
\label{fig:adia_vel}
\end{figure*}
 
In contrast to the isolated halo model \citep{luo18},
cosmological collapse is filamentary and we observe the filaments penetrating inside the growing 
DM halos. The adiabatic accretion flow is traced deep inside the parent halo, where the character of 
the flow changes within  $\sim 10^{-3}$\,pc, as shown in Figure\,\ref{fig:adia_proj_end}. We describe
evolution of this flow on scales $\ltorder 10^{-3}$\,pc below using Figures\,\ref{fig:adia_proj_end} 
and \ref{fig:adia_proj}. On 
these scales, the angular momentum becomes more important, the flow acquires a disk-like character and is
fed by a couple of filaments extending from larger scales. The rotational axis of this flow is slightly 
tilted with respect to the $zx$-plane.

We display snapshots of disk evolution in Figure\,\ref{fig:adia_proj}, approximately along its
rotation axis. The cooling has been artificially and exponentially suppressed at optical depths
$\tau\gtorder 1$.
At about $250.264$\,Myr after the Big Bang, the `photosphere' begins to form, and we re-normalize this 
time to $t=0$. Hence, the photosphere appears within an already disky flow. 
By $t\sim 20$\,yr, the disk becomes stratified, both in radial and vertical directions. 

As the disk becomes visible, it experiences azimuthal distortions, and forms an oval which evolves into a
gaseous bar. The amplitude of this bar varies in time, occasionally becoming very strong, and the bar 
drives a pair of open spiral arms --- a sign that angular momentum redistribution continues. The
disk size reaches $\sim (1-2)\times 10^{-4}$\,pc and $\sim (0.5-1)\times 10^{-4}$\,pc, in radius and in height, 
respectively. So this disk is not geometrically-thin.

After $t\sim 40$\,yr, we observe clump formation in the spiral arms, while the central disk and its bar
do not experience fragmentation. \citet{luo18} argued that 
Kelvin-Helmholtz (K-H) shear instability \citep[e.g.,][]{chan61},
rather than Toomre/Jeans instability is responsible for similar fragmentation seen in simulations of
adiabatic inflow into an isolated DM halo. The main argument is that the clumps form in the spiral
arms or shocks, outside the disk. Such a configuration induces shear in the flow, which may therefore be  
subject to K-H 
instability, when the Richardson number, $Ri < 0.25$. This instability will cause
the shock front to `wiggle,' and clumps will form and grow \citep[e.g.,][]{bal88,kim02}.
The gas self-gravity will act as a stabilizer.
 
The gravitational force across the shock front and postshock layer can be estimated from the
value of the non-axisymmetric term in the gravitational potential, perpendicularly to the shock
front. To estimate $Ri$, we assume that the non-axisymmetric force induced by a spiral arm is a fraction
$\beta\sim 0.05$ of the radial potential measured by the centrifugal acceleration, $v_{\rm t}^2/R$,
where $v_{\rm t}$ is the tangential velocity. This value of $\beta$ is typical for spiral arms.
To project this acceleration perpendicularly to the shock front, we use the pitch angle $i$ of the spiral.
Adopting values from the current run which are similar to those in \citet{luo18},
we obtain $Ri\sim 0.1$, confirming that clumps can form as a result of the K-H shear instability.

Initially, a single clump appears, but additional clumps continue to form in the spiral arms.
Most of the clumps spiral in and merge with the central disk, while 2--3 outer clumps acquire 
angular momentum from the bar and move out. These outer clumps grow faster by accretion, especially after 
$t\sim 60$\,yr. 

The gas density profile is shown for a number of representative times (Fig.\,\ref{fig:adia_param}a),
and is centered on the densest cell of the most massive object, i.e., the disk.
Note the sharp increase in the central density, shortly after formation of the photosphere, to $\sim 
10^{-6}\,{\rm g\,cm^{-3}}$. It decreases thereafter, and increases again.  
By $t\sim 90$\,yr, the disk becomes heavily distorted by the outer clump, its geometry is complicated but
remains planar. Two clumps drive spirals and shocks of their own, and by $t\sim 150$\,yr, there are basically
two cores which grow by accretion. There is in excess of $300\,M_\odot$ within the central $10^{-3}$\,pc
at the end of the run, and much of it is found in the diffuse state. the most massive core is 
$\sim 60\,M_\odot$ (Fig.\,\ref{fig:adia_param}b). 

The large-scale filaments that feed the central disky flow are relatively cold --- their temperature is
close to the temperature floor of the atomic gas at $\sim 3,000$\,K. The disk, outside the photosphere 
is cold as well. At the same time, the clumps are warmer, $T\sim 10^4$\,K, with the most massive clump
being the hottest, and so are the spiral arms 
driven by the asymmetric disk. This higher temperature is associated with larger optical depth for the escaping 
radiation, as well with increased compression.

As the gas shocks and enters the photosphere, its temperature rises sharply by almost a decade. By the
end of the run, the central temperature has reached $\sim 3\times 10^4$\,K, when averaged over
spherical shells (Fig.\,\ref{fig:adia_param}c). More careful analysis shows that the central temperature
fluctuates in the range of $T\sim {\rm few}\times 10^4-10^5$\,K.  

After the first appearance of the photosphere at ${\rm few}\times 10^{-6}$\,pc, it expands out
to $\sim 2\times 10^{-4}$\,pc by the end of the run.  
The mass accretion rate drops inside the photosphere from 
$\sim 1\,M_\odot\,{\rm yr^{-1}}$ by 2--3 orders of magnitude, meaning that rotation and gas pressure 
gradients
terminate the inflow below this radius (Fig.\,\ref{fig:adia_param}d), the radiation pressure being much
smaller at this time. We calculate the accretion rate, $\dot M$, by
measuring the mass difference within spherical radii per timestep.

%% FIGURE 6
\begin{figure*}
\centerline{
 \includegraphics[width=0.77\textwidth,angle=0] {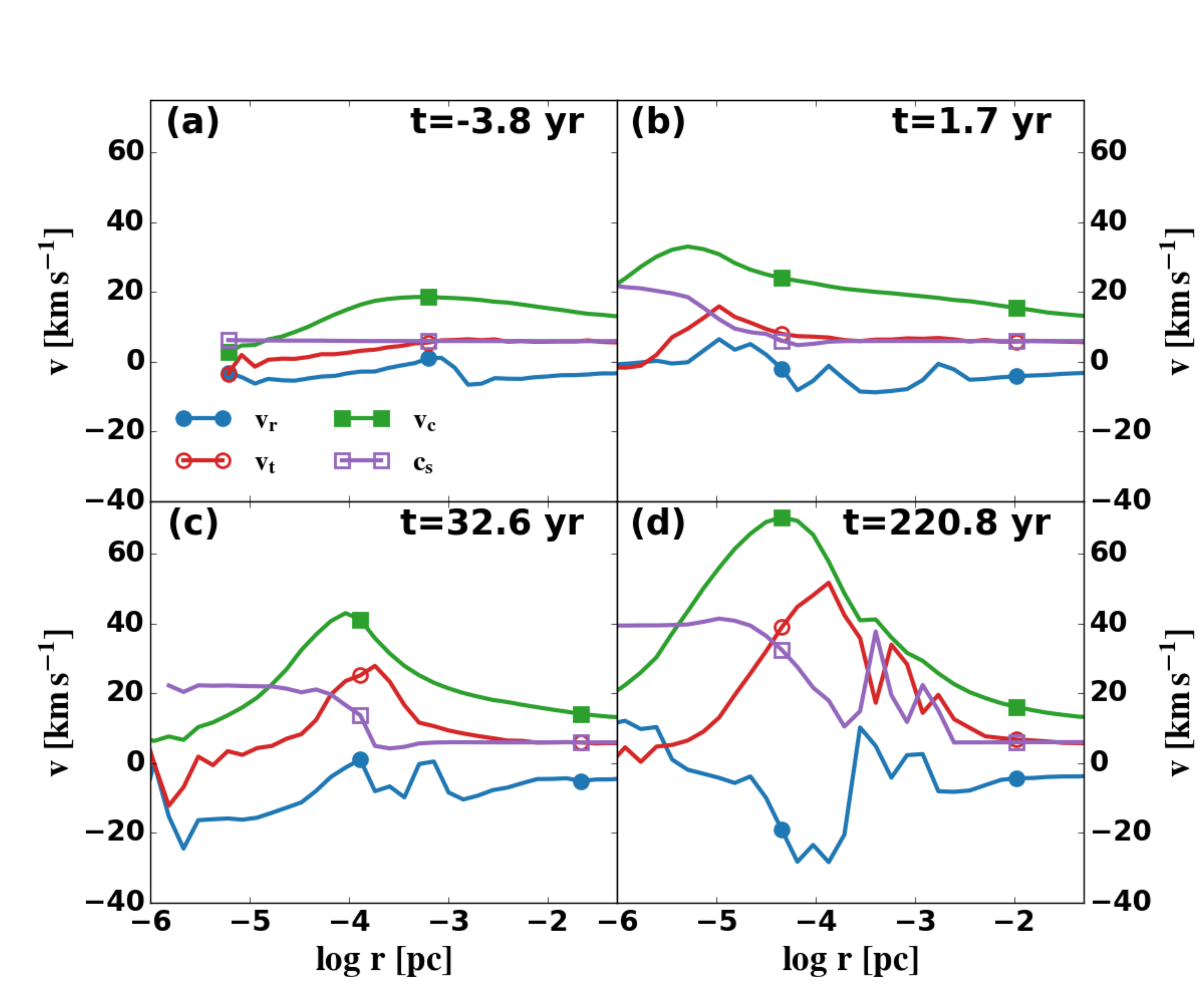}
}
\caption{Adiabatic collapse:  radial profiles of the radial ($v_{\rm r}$), tangential ($v_{\rm t}$), 
and circular ($v_{\rm c}$) velocities,
and sound speed ($c_{\rm s}$) are shown at four representative times, {\it (a)} $t=-3.8$\,yr, 
{\it (b)} $t=1.7$\,yr, {\it (c)} $t=32.6$\,yr, and {\it (d)} $t=220.8$\,yr. Negative values of radial 
velocity represent the inflow. These profiles are centered on the densest cell of the most massive
object, i.e., the disk.
}
\label{fig:adia_vels}
\end{figure*}

The importance of angular momentum in the central region of $\sim 10^{-3}$\,pc is obvious also from
the local velocity field shown in Figure\,\ref{fig:adia_vel} in three projections. The $zx$-projection 
plane shows that the dominant rotation has flattened the object along a plane that is tilted with 
respect to the $zx$-plane. It also confirms that the clumps have an 
intrinsic
temperature higher than the accreting gas, including the disk itself. The clumps show a tendency of
spiraling in and are expected to merge in a couple of orbital periods.

The kinematics of the innermost flow has been quantified in Figure\,\ref{fig:adia_vels}, where radial 
($v_{\rm r}$), tangential ($v_{\rm t})$, and circular ($v_{\rm c})$ velocity profiles, as well as the
sound speed ($c_{\rm s})$, are shown at 4 representative times. The maximum value of $v_{\rm c}(r)$
is increasing with time and moving out until $t\sim 30$\,yr, then stagnates in position but continues
its growth in value --- a sign of accumulating mass and the binary nature of the flow before the two 
clumps merge. The sharp maximum in $v_{\rm r}$ of the last frame in this Figure, at $t\sim 221$\,yrs,
reflects the instantaneous radial velocity of
the neighboring clump on an elliptical trajectory.  This, is also true for the tangential velocity,
which is the result of a complex flow driven by two massive clumps, and is nearly canceled out
in the region between them. Note that these clumps have parallel spins.

\begin{figure}
\centerline{
 \includegraphics[width=0.48\textwidth,angle=0] {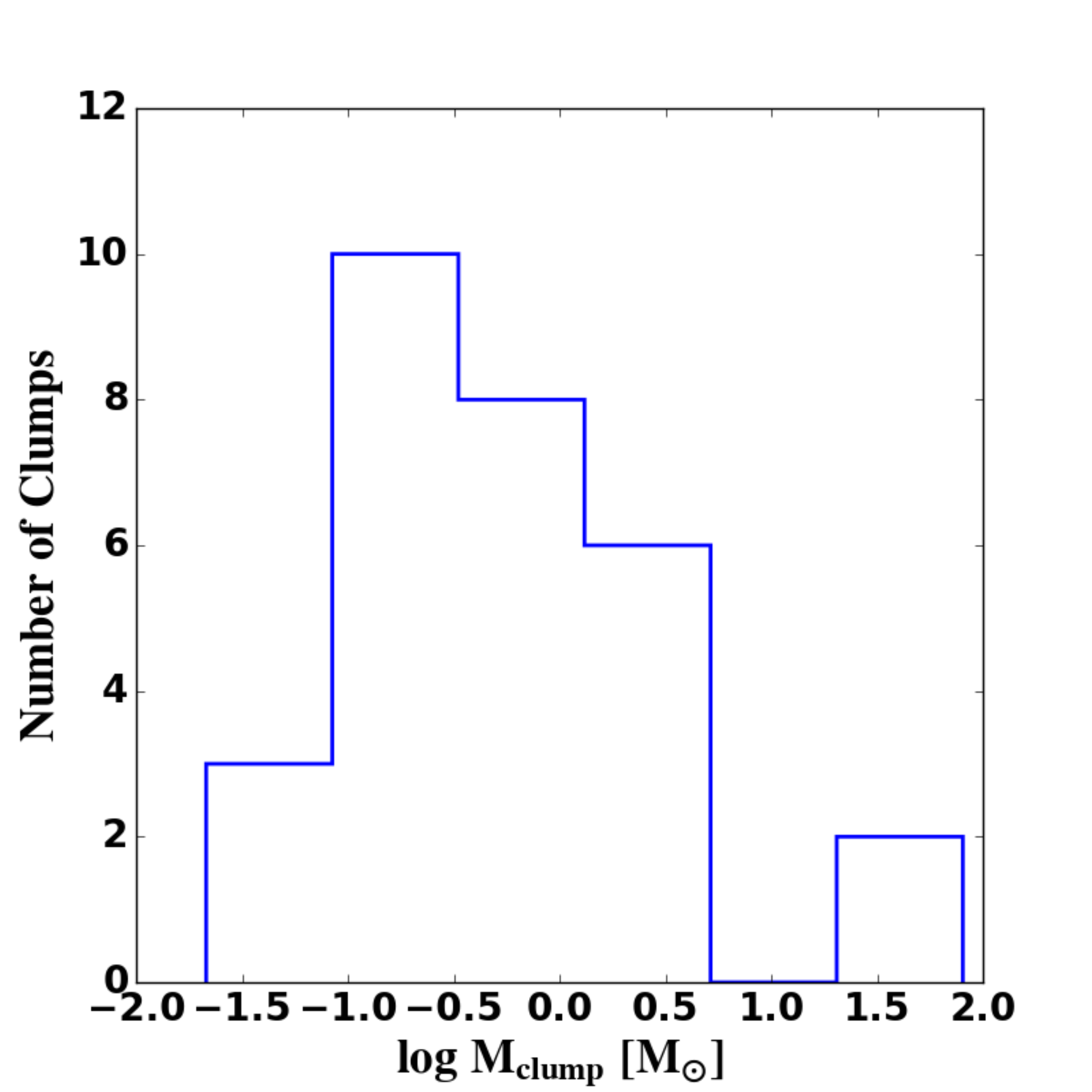}
} 
\caption{Adiabatic collapse: distribution of the clump masses at the end of the simulation,
$t=220.8$\,yr.
}
\label{fig:adia_fragm}
\end{figure}

The distribution of the clump masses at the end of the run is given in Figure\,\ref{fig:adia_fragm}. 
There are two massive clumps
in the $20-90\,M_\odot$ bin. The distribution peaks between $\sim 0.1-0.3\,M_\odot$.

To summarize the adiabatic run, the final configuration consists of a rotationally supported, 
geometrically-thick disk. Being asymmetric, it drives a pair of
spiral shocks at larger radii. The associated fragmentation is limited to the shocked material and 
appears to
be the result of Kelvin-Helmholtz shear instability and not Toomre gravitational instability.
Importantly, the disky character of the inner flow is only perturbed by the clumps but not destroyed. 
The forming clumps constitute a transient phenomenon. The central mass concentration is of the order 
of $\sim 300\,M_\odot$ within the central $10^{-3}$\,pc.

An important question is whether the adiabatic approximation and the resulting disky flow provide an
adequate description for the innermost flow by the FLD approximation, as well. This issue, and the 
comparison between isolated and cosmological runs, are addressed in
Sections\,\ref{sec:FLDresults} and \ref{sec:discuss}.

%% FIGURE 7

%%%%%%%%%%%%%%%%%%%%%%%%%%%%%%%%%%
\section{Results: FLD flow}
\label{sec:FLDresults}
%%%%%%%%%%%%%%%%%%%%%%%%%%%%%%%%%% 

The FLD run does not differ from the adiabatic run at early time and on large spatial scales. We, 
therefore, limit our discussion to the central region, $R\ltorder 10^{-2}$\,pc. The filamentary 
inflow extends to $\sim 10^{-4}$\,pc. 
The inflow is channeled along the main filament, and the outside material joins the filament
after experiencing an oblique shock on its surface. The photosphere forms at $\sim 250.264$\,Myr 
after the Big Bang, which is very similar to that in the adiabatic run. We define this moment as 
$t=0$, which is used in the subsequent evolution.

%%%%%%%%%%%%%%%%%%%%%%%%%%%%%%%%%%%%%%%%%%%%%%%%%%%%%
\subsection{Determining the position of the photosphere}
\label{sec:photo}
%%%%%%%%%%%%%%%%%%%%%%%%%%%%%%%%%%%%%%%%%%%%%%%%%%%%%

For the FLD runs, the position of the photosphere is calculated using a number of alternative options, 
following \citet{luo18}.
First, we trace rays away from the densest cell in the growing core to a distance of 1\,pc, then 
integrate inwards along the ray to the point where $\tau=1$, using the
Planck mean opacity coefficient, $\kappa_{\rm P}$. We use 9,000 rays
equally spaced on a spherical surface. The position of the  photosphere is determined for each ray 
separately. It has no particular symmetry and its shape evolves each timestep. 

Second, the position of the photosphere is calculated using the flux limiter 
(Section\,\ref{sec:rad-transf}), as a trace of the optically-thick region and shown in 
Figure\,\ref{fig:adia_proj}. We assume that the photosphere lies at  $\lambda_{\rm F}=0.33$, where 
$\lambda_{\rm F}$
decreases sharply with radius. LTE has been assumed for the optically-thick part of the flow.

%% FIGURE 8
\begin{figure}
\centerline{
 \includegraphics[width=0.4\textwidth,angle=0] {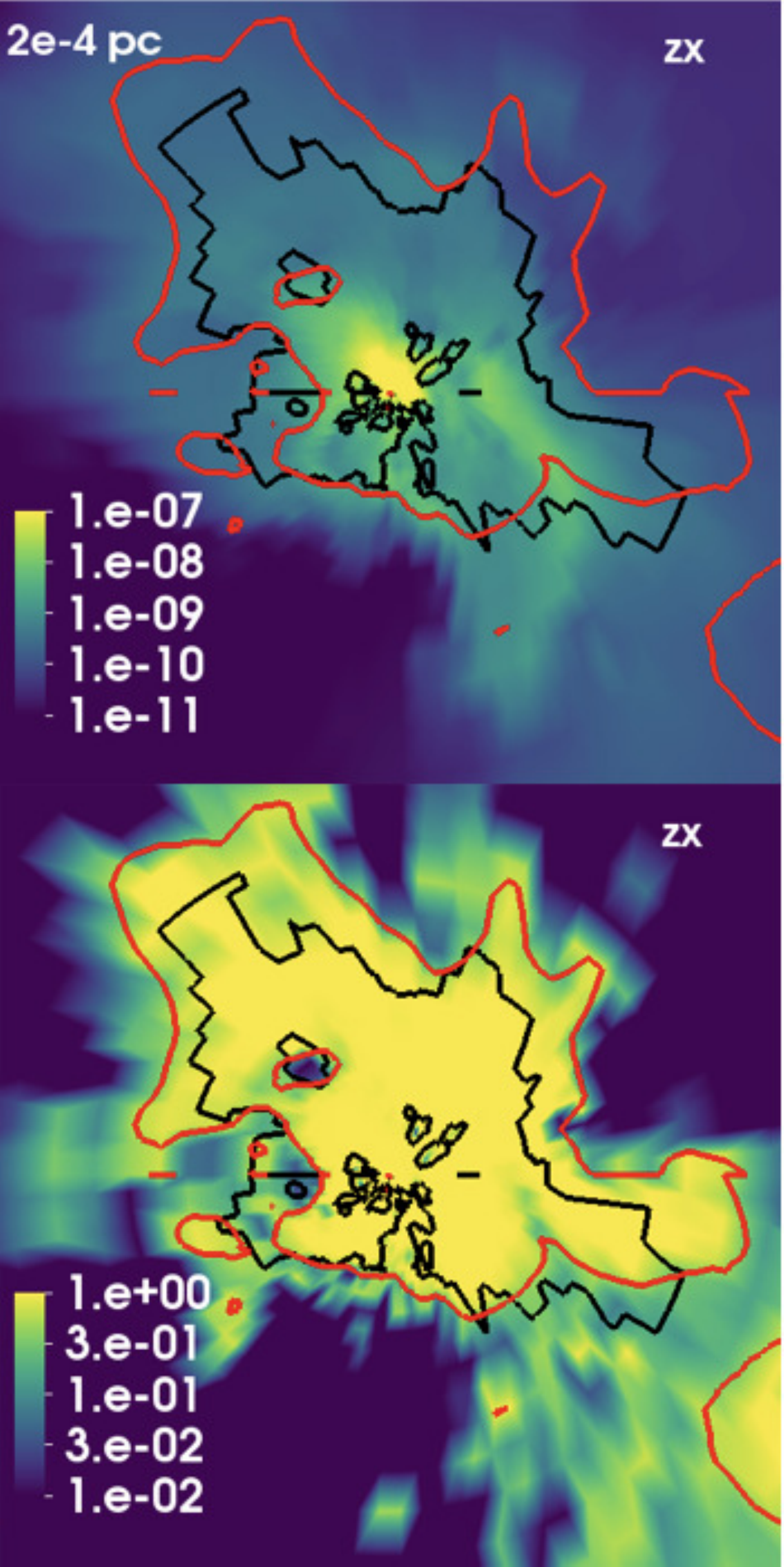}
}
\caption{FLD flow: the flux limiter, $\lambda_{\rm F}=0.33$, contour (black line) and ionization 
fraction of 50\% contour
(red line) superposed on the density slice map (top frame), and ionization slice map (bottom frame)
at $t=37.4$\,yr. The
color palettes are given for each frame. Note the anisotropic ionization map and the generally
neutral gas outside the photosphere which lies immediately outside the contours, meaning that the
Stromgren sphere size is small and the accreting gas is neutral. Each frame is $2\times 10^{-4}$\,pc 
on the side.
}
\label{fig:ioniz}
\end{figure}

The surface $\tau=1$ is slightly offset outwards from the surface of $\lambda_{\rm F}=0.33$. The main 
problem in calculating the 
optical depth is that the ray can intersect the fixed $\tau$ surface a number of times, the 
resulting photospheric radius is often overestimated and we avoid using this method here. We plot the surface 
of constant ionization fraction at the level of 50\% in Figure\,\ref{fig:ioniz}. This surface 
follows the $\lambda_{\rm F}=0.33$ surface sufficiently closely, 
and lies just outside it. Hence both of these latter surfaces can be used for the purpose of 
determining the photospheric shape and radius, which lies between them. 

\begin{figure*}
\begin{center}
\includegraphics[width=0.97\textwidth,angle=0] {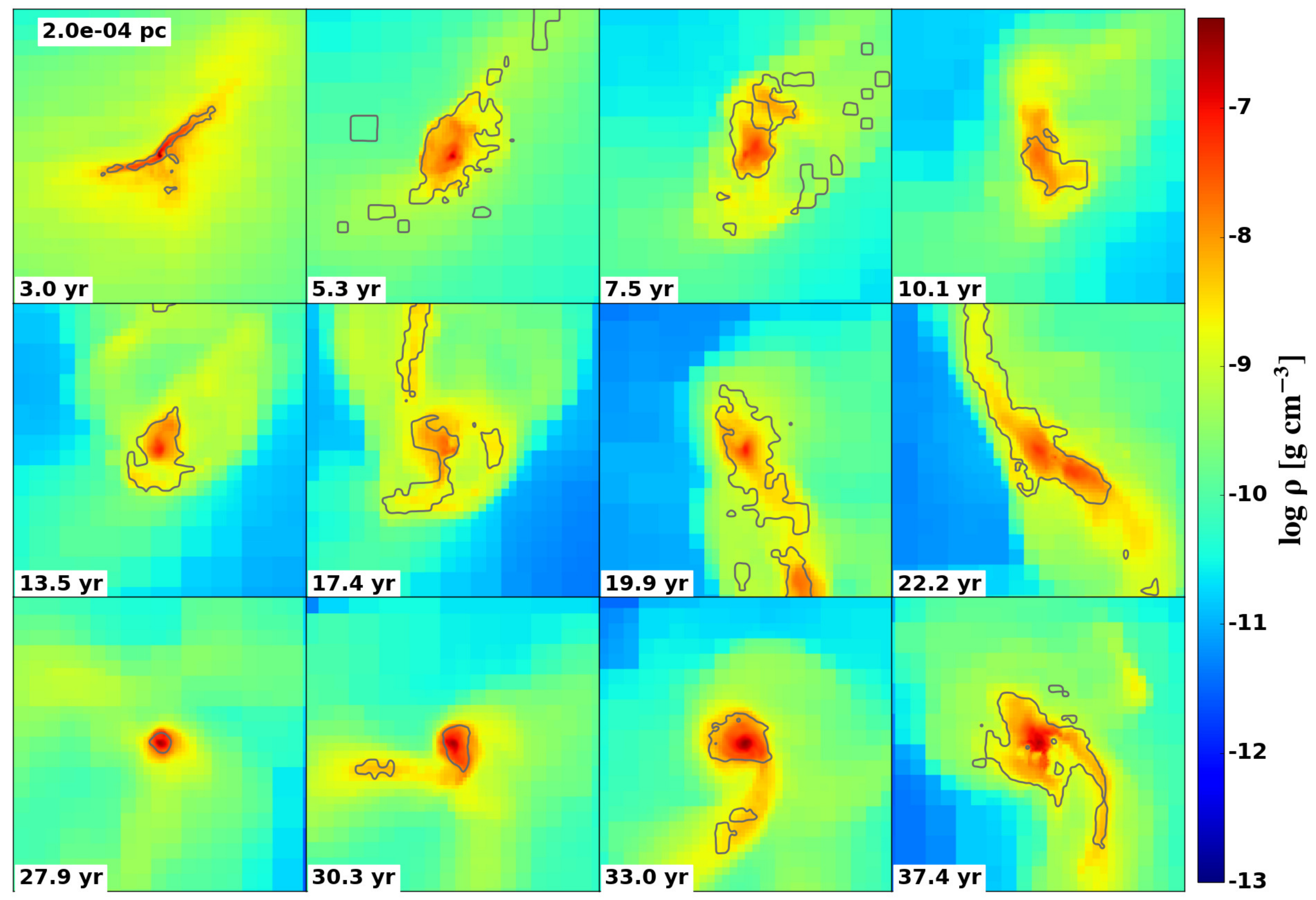}
\caption{FLD collapse: Evolution of the central core and its photosphere, superposed on the 
projected density map. The photosphere is defined using the flux limiter value (\S\,\ref{sec:rad-transf}) 
and is given by the thin solid lines. The color palette provides 
the mass density of the gas. The size of each frame is $2\times 10^{-4}$\,pc.  Note that the field of
view in this Figure is a factor of 10 smaller than shown in Figure\,\ref{fig:adia_proj} for the 
adiabatic run. 
}
\label{FLD-evol2Dphoto}
\end{center}
\end{figure*}

The evolution of the photosphere has been traced in Figure\,\ref{FLD-evol2Dphoto}. Its position has 
been calculated using the flux limiter. Its shape differs from 
averaging in spherical shells, and we avoid using the latter, except in Section\,\ref{sec:lumin}. 

%% FIGURE 9
%%

%%%%%%%%%%%%%%%%%%%%%%%%%%%%%%%%%%%%%%%%%%%%%
\subsection{Central core evolution}
\label{sec:FLDevol}
%%%%%%%%%%%%%%%%%%%%%%%%%%%%%%%%%%%%%%%%%%%%%%

The overall evolution proceeds as follows. The main filament is sheared by rotation on scales 
of $\sim 10^{-5}$\,pc. The mass starts to accumulate where the inflow from the opposite sides of 
the filament encounters itself. As the mass accumulates in the center, the filament itself thickens
and continues to be sheared in the $xz$-plane, i.e., the rotation axis of the flow on this spatial scale
is nearly aligned with the $y$-axis, in agreement with the adiabatic run. The central core of 
$\sim 10^{-5}$\,pc becomes visible 
immediately. The temperatures of the filament and of the core appear to be $\sim (1-2)\times 10^4$\,K,
above that of the surrounding gas, which remains near the floor temperature of the atomic gas.
The shear, which originates in the rotation of the flow, wraps the filament around the core.
The core stays at its original temperature, while the gas around it becomes slightly hotter.
A few solar masses of gas are found within the photosphere by $t\sim 5$\,yr (Fig.\,\ref{FLD-evol2Dphoto}).

\begin{figure*}
\begin{center}
\includegraphics[width=0.97\textwidth,angle=0] {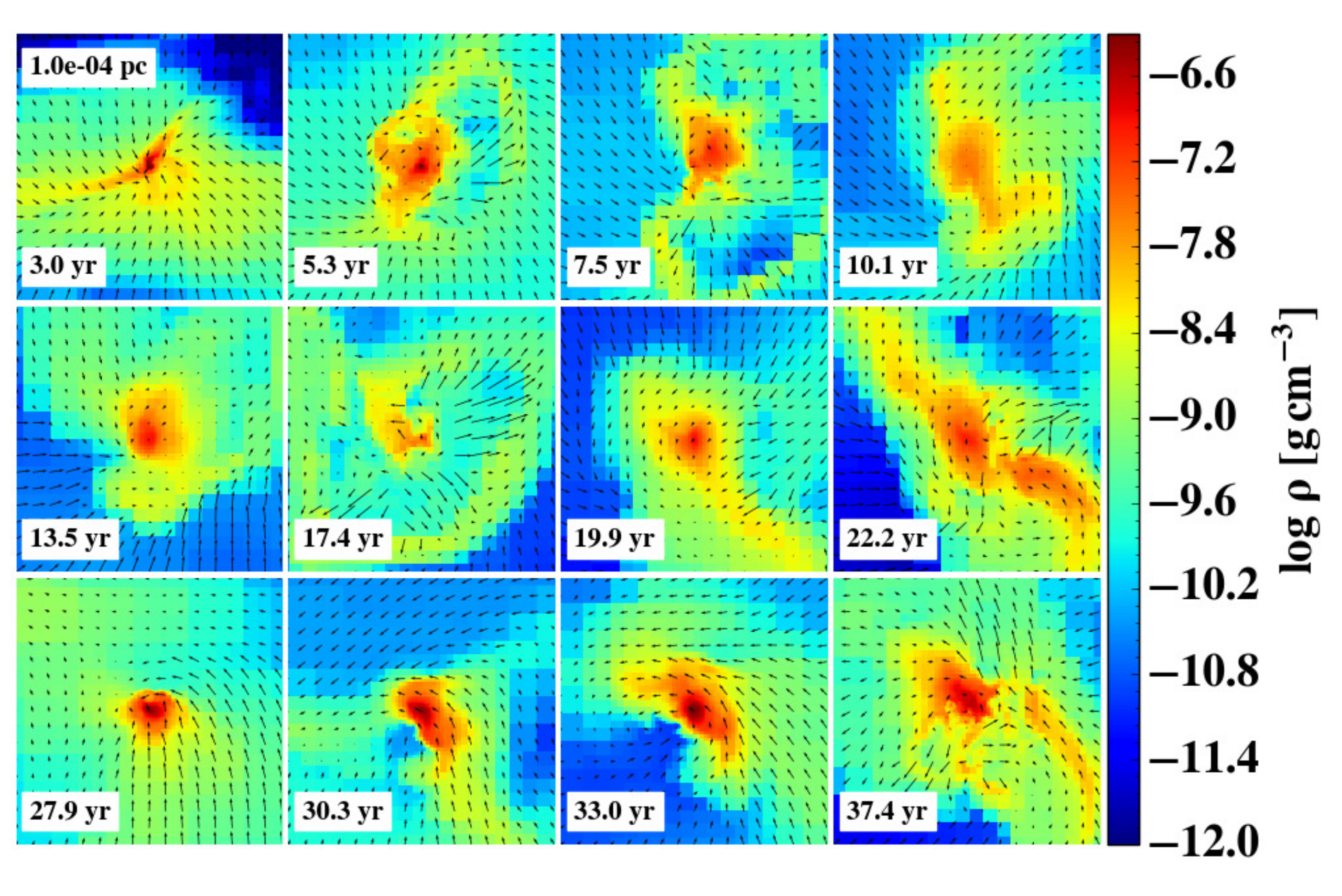}
\caption{FLD collapse: evolution of the velocity vector field of an inner collapsing flow 
superposed onto the density slice map at various representative times. The color palette provides 
the mass density of the gas. Overplotted arrows represent the velocity field and their size is 
proportional to the velocity magnitude. Note the alternate phases of inflow and outflow, and their 
coexistence on various snapshots. The size of each frame is $10^{-4}$\,pc, twice 
smaller than in Figure\,\ref{FLD-evol2Dphoto}.  
}
\label{FLD-evol2Dvel}
\end{center}
\end{figure*}

In a short time of $\sim 4$\,yr after the formation of the photosphere, the first outflow 
develops and is driven by the gas pressure gradients, with some contribution from the radiation 
force which becomes dominant occasionally (Fig.\,\ref{FLD-evol2Dvel}). The outflow is not 
symmetric with respect to the core, nor is the temperature of the gas surrounding it. The solid angle 
of the outflow
is about 1\,steradian. The resulting hot bubble expands rapidly, driving a dense shell, reaching $\sim 
2\times 10^{-4}$\,pc in a few years, where it stagnates. The associated velocity of expansion 
is $\sim 50\,{\rm km\,s^{-1}}$, and hence is supersonic with respect to the ambient gas. Indeed, 
we observe an expanding shock wave. 

%% FIGURE 10

Even with the outflow present, the inflow continues unabated along the filaments. With time, the 
outflow region
starts to envelop the core, still asymmetry remains strong. This affects the filamentary inflow,
which becomes progressively cut off from the core, at least on the side of the outflow. We observe its 
effect on the mass accretion across the photosphere, (see Section\,\ref{sec:lumin}). At 
$t\sim 7-9$\,yr, the mass accretion 
rate dives by about an order of magnitude, and the photospheric radius shrinks visibly.
The mass of the core decreases as well. The temperature of the hot expanding bubble is
$T\sim 10^5$\,K.

As the outflow envelops the central core and evacuates material in its vicinity, we observe 
that the actual shape of the core resembles that of a triaxial ellipsoid, i.e., it is barlike, and
tumbles. During $t\sim 8-11$\,yr, the core appears to be cut off completely from the feeding filaments.
As the outflow ceases by $t\sim 10$\,yr, the accretion resumes, as observed also in velocity maps.
By this time, much of the core mass is `eaten away,' but it resumes its growth. 
This pattern of evolution is followed by another round of outflow, which `eats up' the core 
visibly. 

By $t\sim 12$\,yr, the central structure has lost its disky appearance completely, while the core
grows. Interestingly, the shell driven by the outflow increases its surface density and forms
a mass accumulation by $t\sim 20$\,yr, which behaves like a fragment. Figure\,\ref{FLD-evol2Dphoto}
shows this fragment at $t\sim 22.2$\,yr, already within the photosphere, falling inwards and dissolving 
within the central core shortly thereafter.

The time periods of recurrent outflows appear to come and go, e.g., at $t\sim 17$\,yr, $\sim 22$\,yr,
and especially at $\sim 24-25$\,yr (not shown here), then after $\sim 29$\,yr (seen only in other 
projections), and after $t\sim 36$\,yr.
These outflows have a profound effect on the growth of the mass in the central region. The final snapshot
of the central core in the FLD evolution is shown in Figure\,\ref{FLD-evol2Dphoto} at $t\sim 37.4$\,yr
after the formation of the photosphere.

\begin{figure*}
\begin{center}
\includegraphics[width=0.72\textwidth,angle=0] {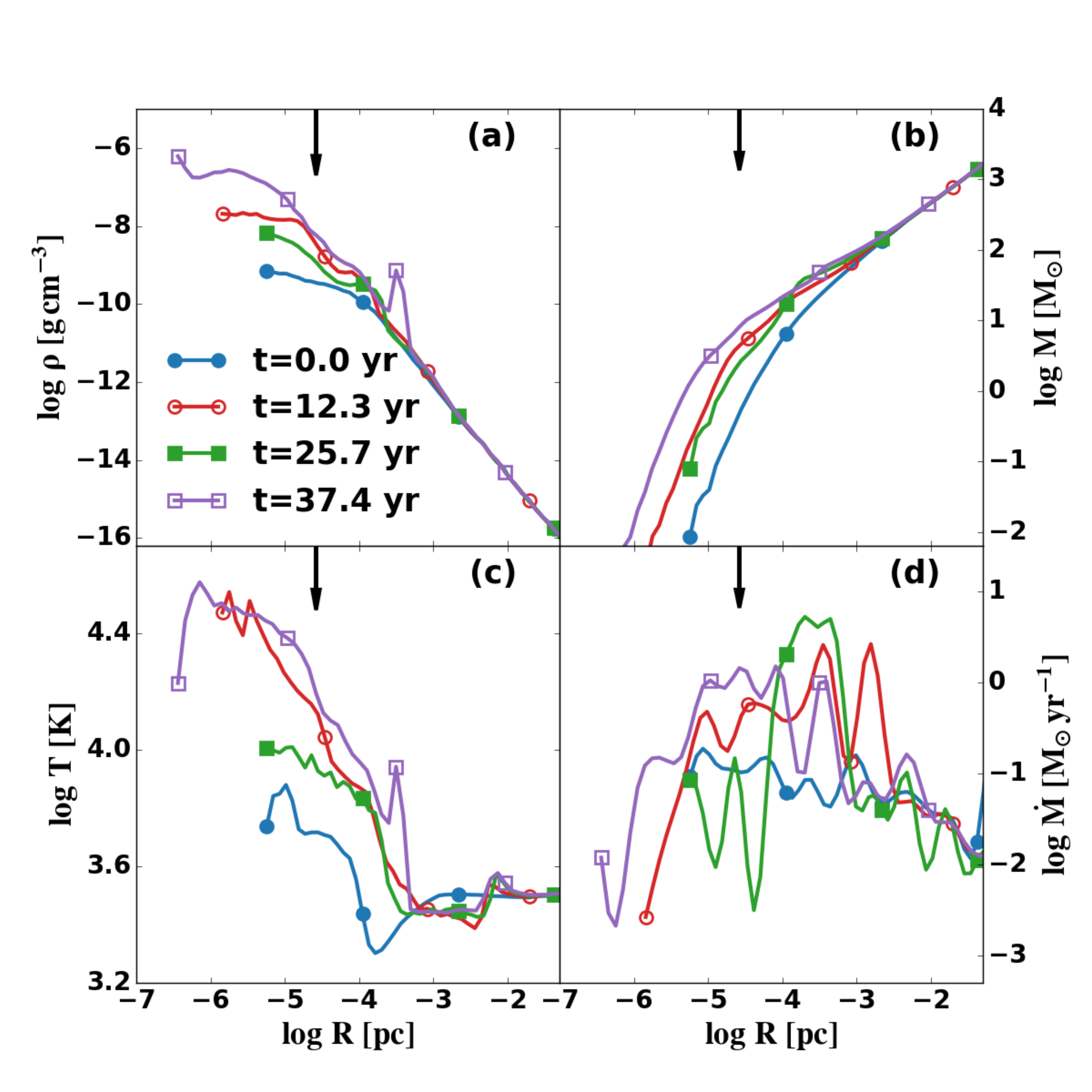}
\caption{FLD collapse: radial profiles of the {\it (a)} gas density, {\it (b)} enclosed gas mass, 
{\it (c)} gas temperature, and {\it (d)} accretion rate at different times. The vertical arrow 
shows the final position of the photospheric radius. The photospheric temperature 
fluctuates but, on average, increases monotonically.
}
\label{FLD-param}
\end{center}
\end{figure*}

The gas density radial profile is shown in Figure\,\ref{FLD-param}a. The central density has
increased substantially, and reached $\sim 10^{-6}\,{\rm g\,cm^{-3}}$ at the end of the run.
The density peak at $R\sim 3\times 10^{-4}$\,pc, visible at $t\sim 37.4$\,yr, is the mass accumulation
at the position of the standing shock of the stagnating bubble. The density profile of the flow departs 
from the $r^{-2}$ law within the photosphere by leveling off. 

Figure\,\ref{FLD-evol2Dphoto} shows that the growth of the central core is not monotonic,
and the shape of the photosphere is heavily affected by both inflow and outflow. 
Within the photosphere, about $10\,M_\odot$ have accumulated by the end of the run, and in excess of 
$100\,M_\odot$ are found inside $10^{-3}$\,pc (Fig.\,\ref{FLD-param}b). The photospheric temperature is
about $3\times 10^4$\,K by the end, up sharply from before the region became optically thick
(Fig.\,\ref{FLD-param}c). The mass accretion rate fluctuates, as we shall discuss later, 
and at the end of the run is just below $1\,M_\odot\,{\rm yr^{-1}}$ (Fig.\,\ref{FLD-param}d).
It drops sharply inside the photospheric radius, but also experiences a local minimum around
$R\sim 10^{-2}-10^{-1}$\,pc, due to the increased importance of the angular momentum there. This outer
minimum in ${\dot M}$ is especially pronounced in isolated models, which have more axisymmetric DM 
halos, compared to cosmological models \citep{choi13,choi15}. 

%%
%% FIGURE 11
%%
%
%% FIGURE 12
\begin{figure*}
\begin{center}
\includegraphics[width=0.72\textwidth,angle=0] {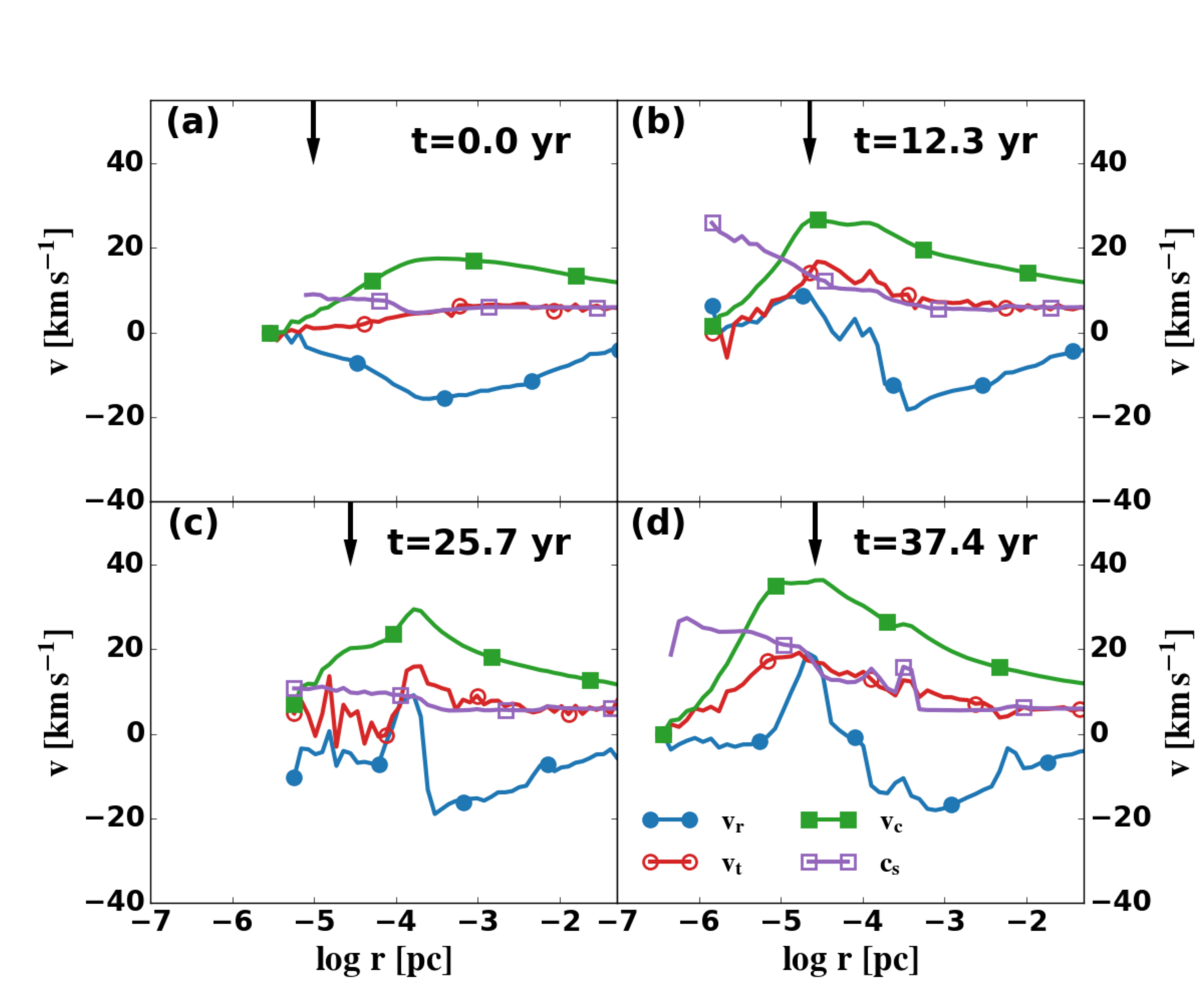}
\caption{FLD collapse: radial profiles of the radial, tangential and  circular velocities and the 
sound speed shown at four representative times. Negative radial velocities correspond to inflow. The 
arrows in each panel 
denotes the position of the photospheric radius calculated based on the averaging over spherical shells, 
except that the tangential velocity averaged on cylindrical shells. The variability arises from  
the appearance of outflows which disrupt the accretion flow temporarily. 
}
\label{FLD-vel_prof}
\end{center}
\end{figure*}

The characteristic velocity profiles of the collapsing flow are given in Figure\,\ref{FLD-vel_prof} at four 
representative times. Circular velocity provides a measure of the radial mass distribution. We observe that 
its maximum value does not increase monotonically --- another signature of alternating inflows and outflows
which affect the central mass accumulation substantially. Radial velocity profiles reflect the same tendency,
e.g., at $t\sim 25.7$\,yr and 37.4\,yr, when the inflow at large radii is reversed at $R\sim 10^{-4}$\,pc 
and $\sim 10^{-5}$\,pc, respectively, and experiences an
outflow. Averaging in spherical shells neglects that both inflow and outflow proceed along particular
directions and often coexist at the same radius. The tangential velocity reaches its maximum around the 
photosphere, but it must be remembered that it is averaged in cylindrical shells and that
counterstreams are present.

%% FIGURE 13
\begin{figure*}
\begin{center}
\includegraphics[width=0.77\textwidth,angle=0] {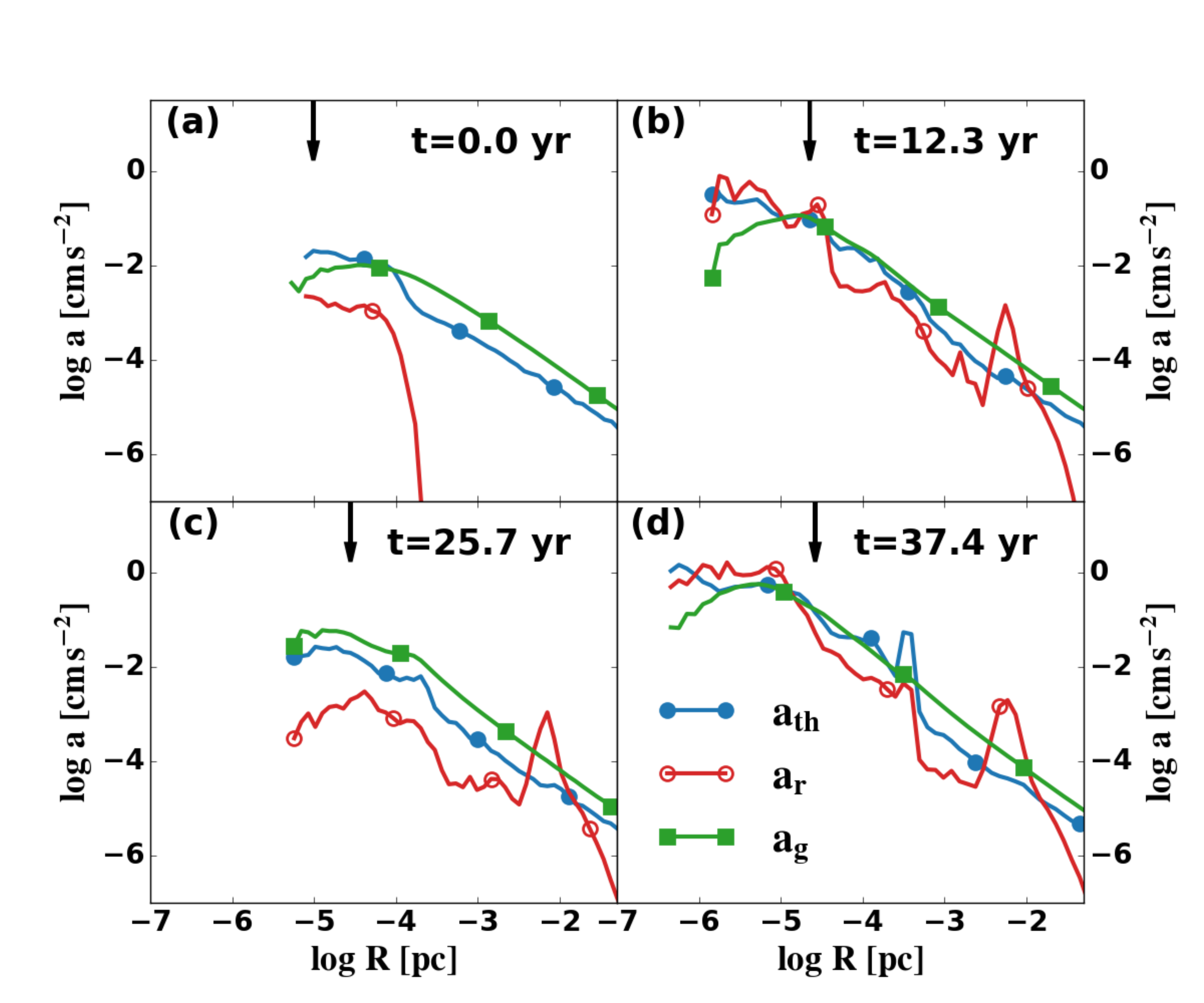}
\caption{FLD collapse: radial profiles of the different types of acceleration: acceleration 
due to the gradient of gas thermal pressure, $a_{\rm th}$, radiative acceleration, $a_{\rm r}$, and 
gravitational acceleration, $a_{\rm g}$, 
presented at four representative times. The arrow in each panel denotes the photospheric radius. 
$t=0$ represents the appearance of the photosphere in the simulation.
}
\label{FLD:accel}
\end{center}
\end{figure*}

Next, we quantify the effect of various forces on the gas, namely, gravity, hydrodynamical force
and radiation force (Figure\,\ref{FLD:accel}). During the formation of the photosphere, at $t=0$, the
radiation force is less important by a factor of a few than the other two forces, up to about a decade 
in radius outside the photosphere. At larger radii the radiation force is completely negligible. Gravity
dominates outside the photosphere, but inside gravity is balanced by radiation and hydrodynamical forces
and some angular momentum. At subsequent times, the radiation force exceeds the hydro force from time
to time, especially around the photosphere and at larger radii. The radial profiles of radiation force
oscillate widely and correlate with the outflow periods. Towards the end of the run the radiation force 
dominates nearly everywhere inside the photosphere.

%%
%%FIGURE 14
\begin{figure*}
\begin{center}
\includegraphics[width=0.77\textwidth,angle=0] {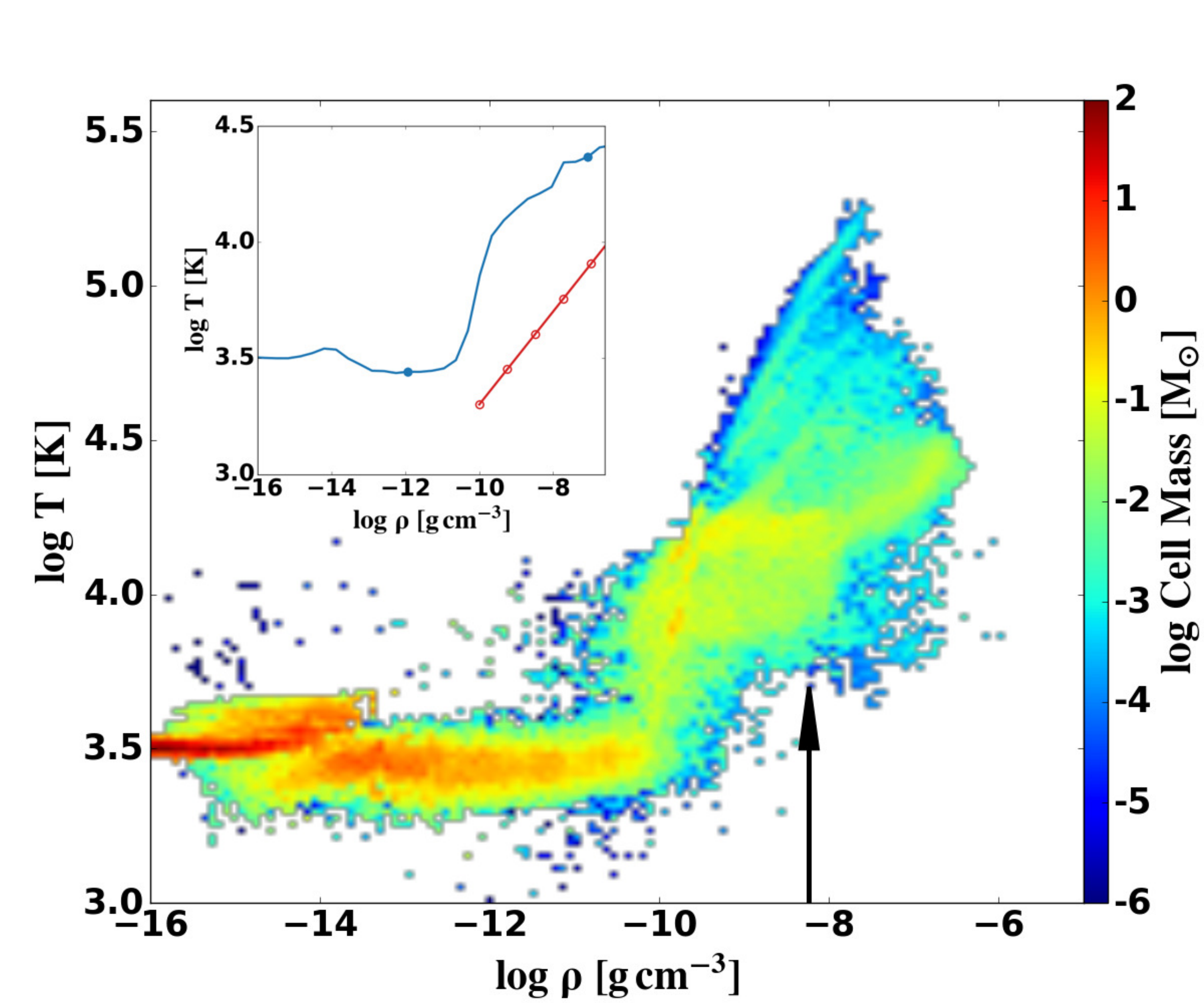}
\caption{FLD collapse: gas temperature as a function of the gas density, at the end 
of the simulation,  $t=37.4$\,yr. The color corresponds to the total mass of all cells for the 
specific  temperature and density. The upward-directed branch reflects the temperature rise due
to an increased opacity in the flow. {\it Inset:} displays the mass-weighted average profile. The red 
line shows a $T\propto\rho^{0.2}$ profile for comparison. Note that adiabatic flow should have
$T\propto \rho^{2/3}$}.
\label{FLD:temp_den}
\end{center}
\end{figure*}

The thermodynamic state of the collapsing gas can be described by averaging properties in spherical 
shells, but this is not always representative of the actual evolution. Therefore, we show the correlation
between the gas temperature and its density using mass averages and displaying all grid cells at the
end of the run (Figure\,\ref{FLD:temp_den}). The gas appears nearly isothermal near the cooling
floor of the atomic gas, for $\rho\ltorder
10^{-12}\,{\rm g\,cm^{-3}}$. Above this density, the temperature distribution for a given density
broadens due to the increasing opacity,
and exhibits a upward-pointing `hot leg' distribution, roughly following $\rho^{0.2}$ (see the inset in 
Figure\,\ref{FLD:temp_den}), due to the gas becoming optically thick. The
important point here is that the temperature is rising slower than in the adiabatic case, where one would
expect $T\propto \rho^{2/3}$. 

To summarize the results of modeling the FLD flow, the character and evolution of this flow differ
fundamentally from the adiabatic flow presented in the previous section and elsewhere in the literature.
The central object which forms in the FLD flow is not dominated by rotation.
After the initial stage, the radiation force determines its dynamics in tandem with gas pressure
gradients. Consequently, no fragmentation occurs. Instead, strong outflows develop but are contained
in the central $\sim 10^{-3}$\,pc. These outflows slow down the growth of the central mass accumulation, 
which reaches about $100\,M_\odot$ within this radius. Moreover, these outflows mix with the massive accretion 
flow and transfer angular momentum outwards, lowering the spin of the central object, within its photosphere. 
This characteristic scale of $10^{-3}$\,pc is determined by the ability of the accretion flow to contain
the outflow, and is about a factor of 100 larger than characteric scales for star formation. Comparison of the
FLD flow with the adiabatic flow is discussed in the next section.

%%%%%%%%%%%%%%%%%%%%%%%%%%%%%%%%%%%%%%%%%%%%%%%%%%%
\section{Discussion}
\label{sec:discuss}
%%%%%%%%%%%%%%%%%%%%%%%%%%%%%%%%%%%%%%%%%%%%%%%%%%%

We have followed direct baryonic collapse within DM halos using high-resolution zoom-in cosmological
hydrodynamic simulations. The inclusion of radiative transfer has allowed us to reach the spatial scales 
of $\sim 0.01$\,AU, 
or $\sim 10^{-7}$\,pc, for the first time taking into account the associated physical processes
involving radiative fluxes and forces in optically-thick and partially thick regions.
The radiative  transfer has been performed in the FLD approximation, and LTE
has been assumed for the optically-thick collapsing region. The adiabatic model has been evolved
for comparison.
  
We find that the collapse proceeds in a filamentary way, and is nearly isothermal in the
outer part, down to $\sim 10^{-4}-10^{-3}$\,pc from the center. The gas is 
channeled  along the filaments, with shocks formed by the material joining the filaments. Inside the
optically-thick region, a central object forms in response to the converging flow and is 
delineated by its photosphere, initially $\sim 10^{-6}$\,pc and expanding thereafter.
Reassuringly, it has a similar size as in the isolated collapse \citep{luo18}. Moreover, the adiabatic 
collapse forms a $\sim 10^{-6}$\,pc `photosphere' in the cosmological run (Section\,\ref{sec:adia_results}), 
and $\sim 2\times 10^{-6}$\,pc in the isolated run. 

This core is well resolved throughout the simulations. Its mass $\sim 10\,M_\odot$
within its photosphere is well above the local cell mass of $\sim 10^{-6}\,M_\odot$, and its 
central density is about $10^{-6}\,{\rm g\,cm^{-3}}$, similar to the adiabatic run.  About $100\,M_\odot$ 
have assembled
within the central $10^{-3}$\,pc, and $\sim 3,000\,M_\odot$ within 0.1\,pc --- again, similar to the
isolated halo runs. The adiabatic run has about 3 times larger mass accumulation in a similar time.

While the central object is clearly identified in the FLD simulation, it is not expected and indeed is not
found to be in hydrostatic or rotational equilibrium.  As the flow enters the $\tau > 1$ region, the radial 
velocity drops, but its internal structure is not relaxed and exhibits
streams and turbulent motions. The tangential velocity increases with radius towards the 
photosphere. The angular momentum profile shows only partial, $\sim $10\% rotational support in 
this region at the end of the run. The remaining support is provided by the buildup of thermal and radiation 
pressure gradients. 

Important new ingredients in the FLD model are the ability of the gas to lose and gain its radiation energy
along the radiation energy gradients
and addition of the radiation force. This results in slower than adiabatic rise of temperature with 
density. In the isolated and cosmological models, we find that
both the kinematics and the dynamics of the FLD flow differs from the adiabatic case.

A number of important issues must be resolved in order to understand the differences between the adiabatic 
and FLD runs in particular, and the advanced stage of direct collapse in general. An incomplete
list of issues includes the following:

\begin{itemize}

\item Why are the cores obtained in adiabatic and FLD runs so inflated compared to the protostellar
stage of massive stars discussed in the literature? 

\item Why does angular momentum dominate the central region kinematics in the adiabatic collapse and not 
in the FLD collapse?

\item Why does fragmentation fail to occur in the central region of the FLD model, contrary to that observed in
the adiabatic model?

\end{itemize}

We start by discussing the first question. The forming cores in the adiabatic and FLD runs are not
relaxed, neither thermally nor dynamically. Density and temperature profiles as well as all major acceleration
profiles, i.e., due to gravity, radiation pressure and gas pressure are variable. The reason for this 
is the large 
accretion rate, which exceeds that encountered in `normal' star formation by orders of magnitude, 
including formation of the Pop\,III stars. Moreover, the accretion rate is strongly variable. An important 
signature of being
out of equilibrium is that the maximal central temperature of the gas is $T\ltorder 7\times 10^4$\,K
in the adiabatic run, and  $T\ltorder 4\times 10^4$\,K in the FLD run, when the core reaches a mass
of $\sim 10\,M_\odot$. While this temperature is insufficient to provide for a hydrostatic support
due to the gas pressure, this is enough to provide radiation pressure support, because the
opacity exceeds that of electron scattering opacity by more than an order of magnitude. In the adiabatic 
model, this support 
is provided mostly by the angular momentum, with some contribution from the pressure gradients.   

%%%%%%%%%%%%%%%%%%%%%%%%%%%%%%%%%%%%%%%%%%%%%%%%%%%
\subsection{Outflows and the angular momentum problem}
\label{sec:outflow}
%%%%%%%%%%%%%%%%%%%%%%%%%%%%%%%%%%%%%%%%%%%%%%%%%%%

Next, we deal with the question of angular momentum redistribution in the adiabatic and FLD runs.
The main difference between these runs is the appearance and even dominance of energetic
outflows driven by a combination of radiation and gas pressure gradients. These outflows are supersonic
and drive shocks into the accretion flow. \citet{luo18} observed the complete core dissolution
in the FLD model, and this phenomenon has reappeared a number of times in the cosmological runs, as
can be seen in Figure\,\ref{FLD-evol2Dphoto} at $t\sim 10.1$\,yr, 27.9\,yr, and $30-33$\,yr. In these
cases the the core did not disappear completely, but lost a substantial mass.
Why are the outflows so powerful in the FLD runs?  They are so powerful, indeed, that they prevent the core 
from growing at the full rate provided by the accretion flow. And what is the fate of the expanding gas?

These outflows break out in specific directions. Typically, as they evolve, they tend to envelop the 
photosphere
after some time, becoming quasi-isotropic. The outflows are stopped around $\sim 10^{-4}$\,pc by
the accretion flow, and mix, presumably via a Rayleigh-Taylor instability. Hence,
accreting gas accumulates in a shell at $\sim 10^{-4}-10^{-3}$\,pc, where it is stirred by the
outflows. This phenomenon is important in that it has no known counterpart in star formation,
where stellar winds from massive stars disperse the surrounding gas as well as the star-forming cloud
itself. In direct collapse considered here, the accretion rate is so high that is capable of 
containing the outflow from the central core.

The ultimate driving force behind these outflows is of course the potential energy
of the accretion flow which is converted into kinetic energy and deposited below the
photosphere. This process is $\sim 10^3-10^5$ times more energetic than during formation of massive
stars, e.g., OB type and Pop\,III stars, because the mass accretion flow is smaller by this factor in the latter 
cases compared to a direct collapse within DM halos. Namely, the accretion rate is 
$\sim 10^{-5}-10^{-3}\,M_\odot\,{\rm yr^{-1}}$ for massive stars and $\sim 1\,M_\odot\,{\rm yr^{-1}}$ is
encountered here. We defer analysis and discussion 
about the nature of these winds to a later publication. 
Outflows play an important role in redistributing the angular momentum in the central region, 
and we elaborate on this below.

An important question emerging from comparison of adiabatic and FLD runs is the dominant role of the
angular momentum in adiabatic models and its secondary role in the FLD models. Clearly this difference
appears only for the innermost flow, roughly within the central $\sim 10^{-4}$\,pc. In this region, the 
two flow realizations differ from each other profoundly.
While we do observe some rotation in the FLD flow prior
to the formation of the photosphere, it is marginalized shortly thereafter. Here we attempt to
address this important issue.

%% FIGURE 15
\begin{figure*}
\begin{center}
\includegraphics[width=0.97\textwidth,angle=0] {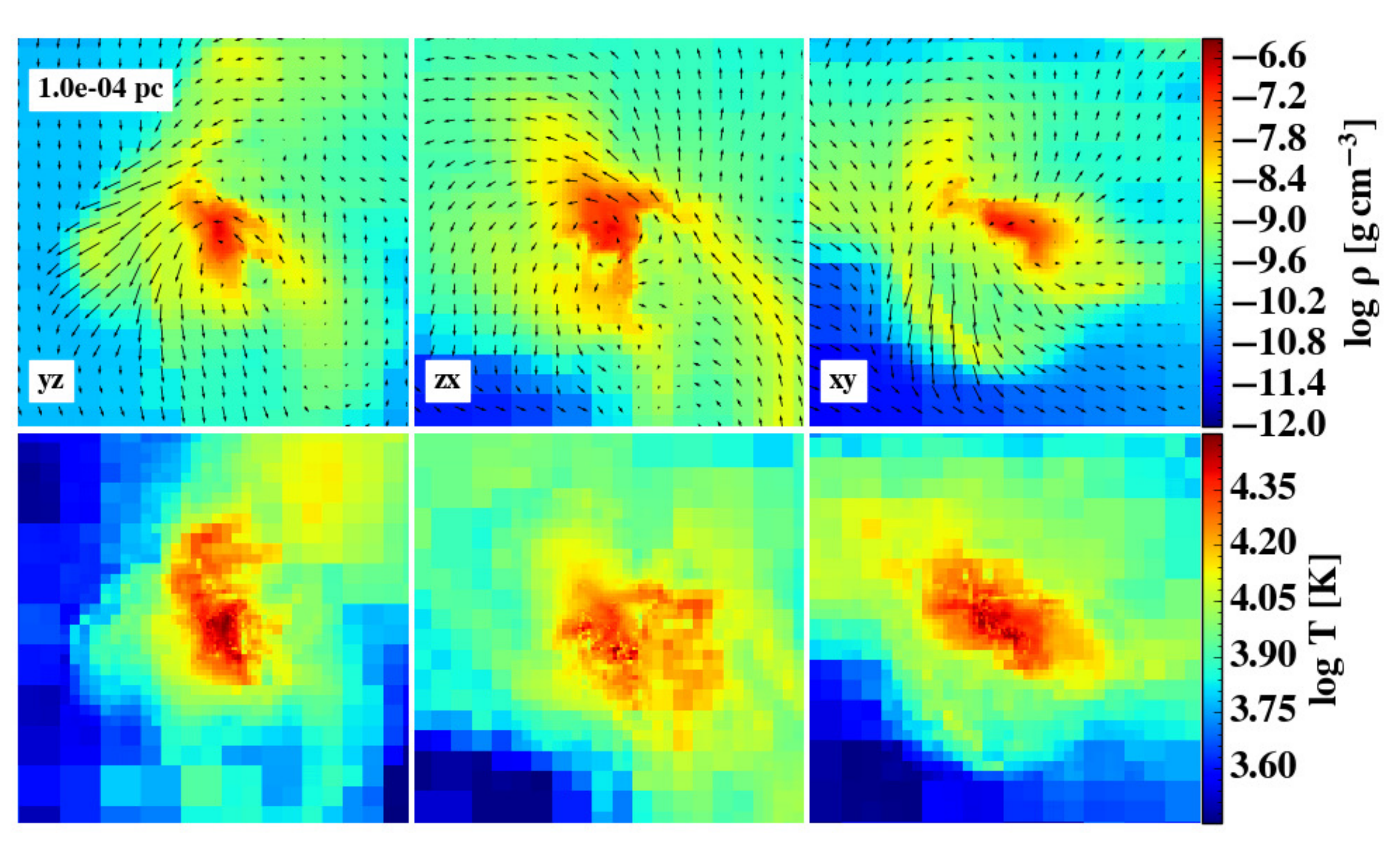}
\caption{FLD collapse: projection snapshots along the three major axes at the end of the simulation,
$t=37.4$\,yr. The velocity field is overplotted on the projected density (top frames), and the temperature 
map is given in the bottom frames. The arrow size is proportional to the velocity value (see 
Figure\,\ref{FLD-vel_prof}), and the color 
palettes are shown on the right margin. The frame size is $10^{-4}$\,pc. Note the anisotropy of the 
outflows.
}
\label{FLD:final_velMAP}
\end{center}
\end{figure*}
%%  
% FIGURE 16
\begin{figure}
\begin{center}
\includegraphics[width=0.45\textwidth,angle=0] {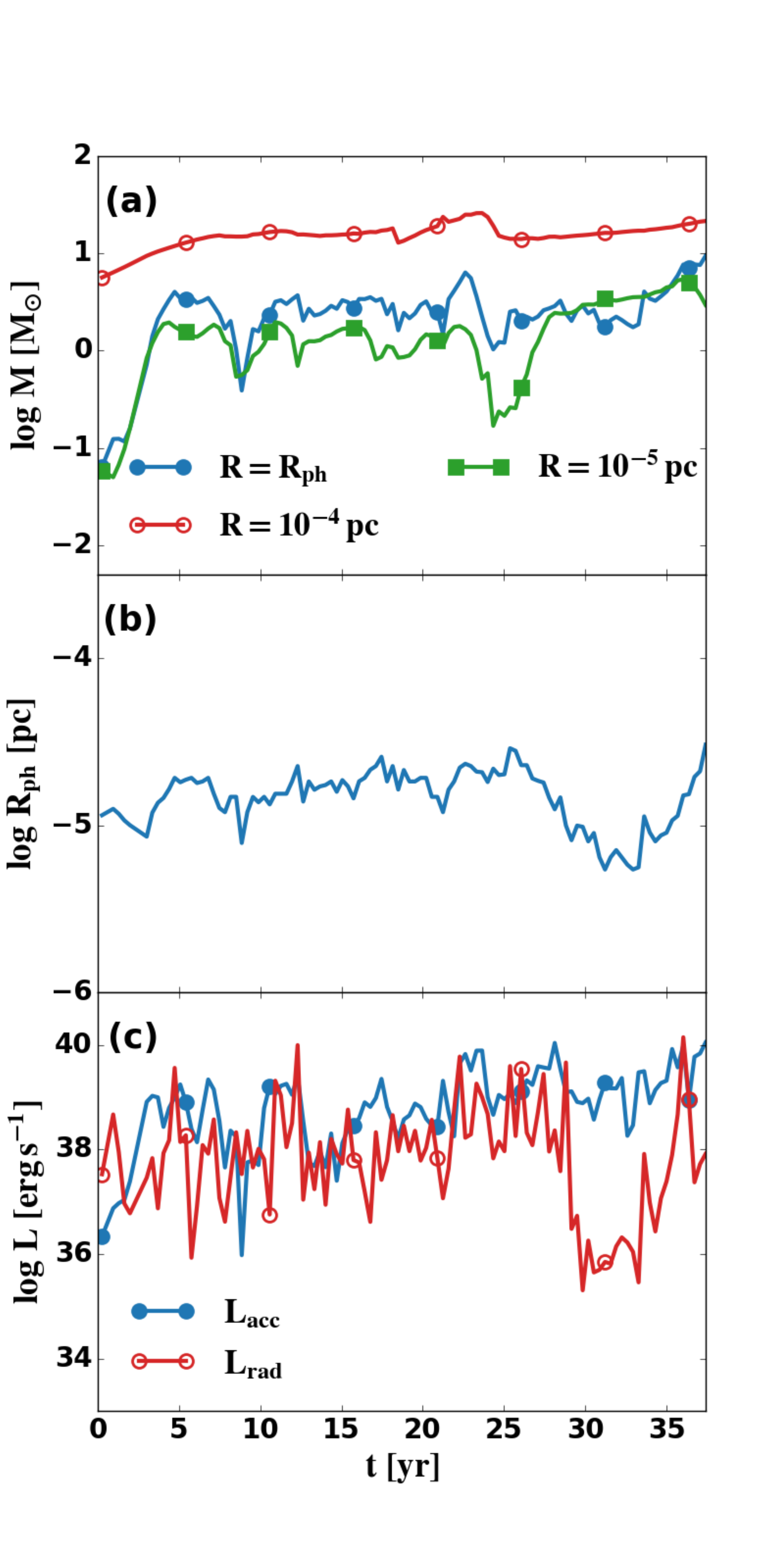}
\caption{FLD collapse: time evolution of {\it (a)} the mass within the photospheric radius, $R_{\rm ph}$, 
enclosed by the mass within $10^{-5}$\,pc and $10^{-4}$\,pc, respectively, for clarity; {\it (b)} the 
photospheric radius, $R_{\rm ph}$; and {\it (c)} the 
mass accretion luminosity (line with full circles) and radiative luminosity (line with open circles), 
$L_{\rm acc}$ and $L_{\rm rad}$, respectively.
All quantities have been calculated at the photospheric radius which has been obtained averaging
in the spherical shells. Note, that this method is more approximate than the other methods used in this work,
and, therefore, this radius differs somewhat from those used elsewhere in the text.
}
\label{FLD:lum}
\end{center}
\end{figure}

On larger scales, the angular momentum is redistributed by gravitational torques and induced shocks.
On scales $\sim 10^{-4}-10^{-3}$\,pc of the adiabatic flow, the angular momentum is transferred outward
by the recurrent bar-like perturbation which drives strong spiral shocks (Figure\,3). No such
configuration forms in the FLD case, despite the initial sheared flow that is present.

As expected, the molecular or ion viscosity has an enormously long timescale and can be neglected. But
the forming core drives strong anisotropic outflows repeatedly (e.g., Figures\,\ref{FLD-evol2Dvel}
and \ref{FLD:final_velMAP}), which extend
to $\sim 10^{-4}$\,pc at $t\sim 17$\,yr (see associated snapshot in Figure\,\ref{FLD-evol2Dvel}), 
and stir the gas within this  region, mixing with the accretion flow, and
concurrently mixing its angular momentum. Note that the specific angular momentum of the accretion flow
is substantially higher than the angular momentum of the outflow. 
Such a turbulent region drives shocks and is capable of transferring angular momentum outwards.

We, therefore, estimate the timescale invoking a turbulent viscosity in order to extract the angular 
momentum from the core and vicinity. The timescale for viscosity to have an effect is $t_{\rm turb}\sim
R_{\rm turb}^2/\nu$, where $\nu\sim l\,v_{\rm turb}$ is kinematic turbulent viscosity, $l$ is the mean free path
for a turbulent cell, $R_{\rm turb}$ is the size of the turbulent region, and $v_{\rm turb}$
is the typical turbulent velocity.  We take $R_{\rm turb}\sim l\sim 10^{-4}$\,pc, $v_{\rm turb}\sim
c_{\rm s}\sim 2\times 10^6\,{\rm cm\,s^{-1}}$ --- all from the FLD runs.

The resulting turbulent viscosity timescale is,

\begin{equation}
t_{\rm turb}\sim 4.8\,\bigg(\frac{l}{10^{-4}\,{\rm pc}}\bigg)\,\bigg(\frac{v_{\rm turb}}
   {20\,{\rm km\,s^{-1}}}\bigg)^{-1}\,{\rm yr},
\end{equation}
meaning that a turbulent flow can redistribute the angular momentum on a timescale that
we observe in the FLD run. This short timescale explains one of the main differences between the adiabatic 
and FLD flows.

%%%%%%%%%%%%%%%%%%%%%%%%%%%%%%%%%%%%%%%%%%%%%%%%%%%
\subsection{Luminosity of the central pre-SMBH object}
\label{sec:lumin}
%%%%%%%%%%%%%%%%%%%%%%%%%%%%%%%%%%%%%%%%%%%%%%%%%%%

One of the differences between the FLD and adiabatic runs is the ability of the former cells to radiate
energy in addition to the expansion and contraction available to the cells in the adiabatic 
approximation. 
We calculate the radiative luminosity of the pre-SMBH object when it forms a photosphere. This is 
done by using Fick's law (Eq.\,\ref{eq:Fick}). The emerging photospheric radiation
luminosity after $t\sim 5$\,yr is $L_{\rm rad}\sim {\rm few}\times 10^{38}-{\rm few}\times 10^{39}\, 
{\rm erg\,s^{-1}}$ --- about the Eddington luminosity for this mass, and exhibits 
strong variability above and below this range, as shown in Figure\,\ref{FLD:lum}c. As the central core 
grows in mass and in size, its radiation luminosity grows in tandem.

The energy source of this radiation comes from the potential energy of the accretion flow, 
$L_{\rm acc}$, that is
converted into kinetic energy and deposited inside the photosphere, {\bf where it is randomized and thermalized.} 
The kinetic luminosity of accretion
can be estimated as $\sim0.5\dot M v_{\rm R}^2\sim10^{38}\,(\dot M/1\,M_\odot\,{\rm yr^{-1}})\,(v_{\rm
R}/20\,{\rm km\,s^{-1}})^2\,{\rm erg\,s^{-1}}$, and varies with time as well. But one 
should not overlook the associated accretion of the gas thermal energy, $\sim 2\pi\rho c_{\rm s}^2\,
R_{\rm ph}^2\,v_{\rm R}\sim 3\times 10^{38}\,(R_{\rm ph}/4\times 10^{-5}\,{\rm pc})^2\,
(v_{\rm R}/20\,{\rm km\,s^{-1}})\,(c_{\rm s}/15\,{\rm
km\,s^{-1}})^2\,{\rm erg\,s^{-1}}$. Here all the values are taken at the end of the FLD run, and 
$R_{\rm ph}$ is the photospheric radius. Hence, unlike accretion on compact objects, the 
contribution of the second term can play an important role in direct collapse.   

Obviously, this estimate indicates that the forming core in direct collapse should be relatively
loosely bound. In other words, the core material rather `levitates' within the photosphere due to
a dominant radiation force -- this is supported by Figure\,\ref{FLD:accel}d. 
This situation is expected to be maintained at least until the core grows 
substantially above the characteristic stellar mass. The multiple 
periods of outflows, which we have observed during the FLD run, confirm this expectation. 

Figure\,\ref{FLD:lum} displays the evolution of the photospheric mass, radius and radiation and 
accretion 
luminosities. Strong variability characterizes the evolution of all of these quantities. This variability
is about a factor of 100 in amplitude for $L_{\rm acc}$ and about 1,000 for $L_{\rm rad}$, superposed 
on steady growth with an average accretion rate of $\sim 0.3\,M_\odot\,{\rm yr^{-1}}$. Hence the rate
of supplied accretion energy varies substantially less than the emerging radiation luminosity.

The photospheric radius, which shows a slow growth after $t\sim 5$\,yr (Fig.\,\ref{FLD:lum}b), exhibits 
a deep minimum around $t\sim 30-33$\,yr --- a consequence of a strong outflow, which appears
to be a response to the peak $L_{\rm rad}\sim 10^{40}\,{\rm erg\,s^{-1}}$ prior to this time. This 
alternating growth and decrease in the mass of the core leads to a complicated behavior of $L_{\rm rad}$ 
and $L_{\rm acc}$ in Figure\,\ref{FLD:lum}c.
There is only a remote correspondence between their oscillations. The reason for this lies
in the ability of the mass accretion flow to deposit and store energy deep within the photosphere.

The typical diffusion time of the photons from the depth of $l\sim 0.1\,R_{\rm ph}$ is 
$t_{\rm diff}\sim l^2/2D\sim 0.2\,(R/4\times 10^{-6}\,{\rm pc})^2$\,yr, where $l$ is the characteristic 
diffusion radius. This timescale has been Fourier analysed in \citet{luo18} for a $1\,M_\odot$ core
to be at $\sim 0.12$\,yr. The same analysis repeated here for Figure\,\ref{FLD:lum}c results in an
identical timescale of $\sim 0.10$\,yr timescale. Accretion
luminosity exhibits a characteristic timescale of $\sim 5$\,yr, compared to 10\,yr in the isolated
case. This timescale is related to that of the accretion flow itself and hence can be affected 
by the outflow feedback.

%%%%%%%%%%%%%%%%%%%%%%%%%%%%%%%%%%%%%%%%%%%%%%%%%%%
\subsection{No fragmentation in the FLD flow}
\label{sec:fragment}
%%%%%%%%%%%%%%%%%%%%%%%%%%%%%%%%%%%%%%%%%%%%%%%%%%%

The kinematics of the FLD flow exhibits no dominant role for angular momentum in the core and its vicinity.
The initial disk-like flow within the central region does not acquire rotational support because it
is capable of transferring its angular momentum via turbulent mixing between the radiation-driven outflow
and accretion inflow. This turbulent flow outside the photosphere and very optically-thick flow
inside the photosphere are not prone to fragmentation. 

On the other hand, the accretion flow on similar spatial scales in the adiabatic flow forms a disk,
partially supported by internal pressure, which also thickens it in the vertical direction. This
disk is subject to non-axisymmetric perturbations, mainly the $m=2$ mode, and drives spiral
shocks (Fig.\,\ref{fig:adia_proj}). We have argued here and in \citet{luo18}, that shocks in
the sheared accretion flow will drive the Kelvin-Helmholtz instability and form clumps. Indeed, these
clumps form in the spiral shocks and not in the disk itself, which supports our argument.
Therefore, we do not agree with the view that the forming clumps are result of gravitational fragmentation
in the disk \citep[e.g.,][]{bec15,bec17}. In fact, the disk is geometrically thick, which damps the
fragmentation exponentially \citep{too64}.

%%%%%%%%%%%%%%%%%%%%%%%%%%%%%%%%%%%%%%%%%%%%%%%%%%%
\section{Conclusions}
\label{sec:conc}
%%%%%%%%%%%%%%%%%%%%%%%%%%%%%%%%%%%%%%%%%%%%%%%%%%%

We have modeled gravitational collapse of a primordial gas within DM halos, including radiative transfer
following the establishment of a photosphere. This corresponds to the most advanced stage of direct collapse
to form seeds of SMBHs at high redshift in a cosmological framework performed so far. Using high-resolution 
zoom-in cosmological simulations, we have compared runs with an adiabatic equation of state with those in 
the flux-limited diffusion (FLD) approximation.

We have observed the formation of central cores surrounded by an irregularly-shaped photosphere, nearly 
simultaneously, 250\,Myr, after the Big Bang, i.e., at $z\sim 15.8$, in both approaches. Yet the properties 
of the cores appear 
to be quite different. A rotationally-dominated core, in the form of a geometrically-thick disk, forms in the 
adiabatic run, supplemented by
smaller fragments forming as a result of Kelvin-Helmholtz (K-H) shear instability in the spiral arms
driven by an asymmetric disk. These fragments 
are transient and eventually merged with the disk, which has a mass of $\sim 100\,M_\odot$. The central mass
concentration achieved at the end of the adiabatic run is  about $300\,M_\odot$ 
assembled within the central $10^{-3}$\,pc, and $\sim 3,000\,M_\odot$ within the central 0.1\,pc. 

The central region of the FLD flow evolves differently. The initially disky
flow within the central $10^{-3}$\,pc quickly loses its angular momentum and an amorphous core 
develops and grows to $\sim 10\,M_\odot$ within a  photosphere of close to $10^{-4}$\,pc. No fragmentation 
is observed to occur because the central region has lost its angular momentum rapidly and the K-H
instability does not operate there. The mass concentration on larger, $\gtorder 10^{-3}$\,pc scales is 
similar to that in the adiabatic flow, but its dynamics is fundamentally different.

The absence of dominant rotational support of the central object in the FLD run, 
is due to the development of massive outflows, triggered by the presence of radiation force and gas pressure 
gradients. These recurrent supersonic outflows are found to drive dense shells of gas by about a factor of 
10 in radius, in essence cutting off the core from the accretion flow. They also are responsible for 
redistributing the angular momentum away from the core.

The core radiation luminosity in the FLD run is of the order of the Eddington luminosity, and highly 
variable, i.e., $L_{\rm rad}\sim 10^{38}-10^{39}\,{\rm erg\,s^{-1}}$. 
The cores in both runs are substantially inflated in comparison to expected protostellar sizes of comparable 
masses. The reason for this is the dominant radiation pressure within the photosphere, which results
in the gas essentially levitating at the Eddington limit.

We confirm our previous result, obtained for direct collapse in isolated halos \citep{luo18}, that
radiation transfer allows the gas in the central structure to cool due to anisotropic density and
thermal gas and radiation gradients, in the presence of an irregularly-shaped photosphere. The
resulting rise of temperature with density is substantially shallower than the adiabatic rise. This result
is in a strong contrast with the adiabatic approximation for the equation of state used currently in the
literature.

%% =============================================================================
\section*{Acknowledgments}
We thank the Enzo and yt support team for help. All analysis has been conducted using yt
\citep{yt}, http://yt-project.org/ and VisIt \citep{chi12}.
We thank Daniel Reynolds for help with the FLD solver, and Kazuyuki Omukai for providing the updated
opacities for a comparison. Discussions with Pengfei Chen, Michael Norman, Kazuyuki 
Omukai, Daniel Reynolds, and Kengo Tomida are 
gratefully acknowledged. This work has been partially supported by the Hubble Theory grant 
HST-AR-14584, and by JSPS KAKENHI grant 16H02163 (to I.S.). 
I.S. and K.N. are grateful for a generous support from the International Joint Research 
Promotion Program at Osaka University. JHW acknowledges support from NSF grant AST-1614333, Hubble 
Theory grants HST-AR-13895 and HST-AR-14326, and NASA grant NNX-17AG23G. MB acknowledges NASA ATP grants 
NNX14AB37G and NNX17AK55G and NSF grant AST-1411879. The STScI is 
operated by the AURA, Inc., under NASA contract NAS5-26555. Numerical simulations have been
performed on XC30 at the Center for Computational Astrophysics, National Astronomical Observatory of Japan,
on the KDK computer system at Research Institute for Sustainable Humanosphere, Kyoto University,
on VCC at the Cybermedia Center at Osaka University, as well as on the DLX Cluster of the University
of Kentucky.

%\bibliographystyle{mn}
%\bibliography{MyRef}

\end{document}